\documentclass[10pt,twocolumn,letterpaper]{article}

\usepackage[applications]{wacv_24/wacv}      

\usepackage{amsmath}
\usepackage{listings}
\usepackage{multirow}
\usepackage{xcolor}
\usepackage{pifont}
\usepackage{xfrac}
\usepackage{makecell}
\usepackage{stfloats}
\usepackage[export]{adjustbox} 

\usepackage{multicol}
\usepackage{booktabs}
\usepackage{url}
\usepackage[bottom,hang,flushmargin]{footmisc}

\newcommand{\xfill}[2][.7ex]{{%
  \dimen0=#2\advance\dimen0 by #1
  \mbox{}\leaders\hrule height 2.9pt depth -#1\hfill%
}}

\usepackage[pagebackref,breaklinks,colorlinks]{hyperref}

\usepackage[capitalize]{cleveref}
\crefname{section}{Sec.}{Secs.}
\Crefname{section}{Section}{Sections}
\Crefname{table}{Tab.}{Tables}
\crefname{table}{Table}{Tabs.}
\crefname{figure}{Figure}{Figs.}

\usepackage{amsmath,amsfonts,bm}

\def\floor#1{\left \lfloor #1 \right  \rfloor}
\def\1{\bm{1}}

\DeclareMathAlphabet{\mathsfit}{\encodingdefault}{\sfdefault}{m}{sl}
\SetMathAlphabet{\mathsfit}{bold}{\encodingdefault}{\sfdefault}{bx}{n}

\newcommand{\lp}{\left(}
\newcommand{\rp}{\right)}
\newcommand{\lb}{\left[}
\newcommand{\rb}{\right]}

\newcommand{\xscalar}{x}

\newcommand{\xvec}{{\bf \xscalar}}
\newcommand{\fvec}{{\bf f}}
\newcommand{\rvec}{{\bf r}}

\newcommand{\yscalar}{y}
\newcommand{\yvec}{{\bf \yscalar}}

\newcommand{\muvec}{\boldsymbol{\mu}}

\newcommand{\sscalar}{s}
\newcommand{\svec}{{\bf \sscalar}}
\newcommand{\uvec}{{\bf u}}
\newcommand{\sigmavec}{{\bf \sigma}}
\newcommand{\rhovec}{{\bf \rho}}

\newcommand{\Real}{\mathbb{R}}

\newcommand{\round}[1]{\ensuremath{\left\lfloor#1\right\rceil}}

\newcommand{\warp}[1]{\operatorname{warp}\lp #1 \rp}

\newcommand{\DenseWarp}[1]{\operatorname{warp}_\text{dense}\lp #1 \rp}
\newcommand{\BlockWarp}[1]{\operatorname{warp}_\text{block}\lp #1 \rp}
\newcommand{\OverlapBlockWarp}[1]{\operatorname{warp}_\text{block-overlap}\lp #1 \rp}

\def\wacvPaperID{2554} 
\def\confName{WACV}
\def\confYear{2024}

\newcommand{\snapdragon}{Snapdragon\textsuperscript{\tiny\textregistered}~}

\newcommand{\cmark}{\textcolor{teal}{\ding{51}}}%
\newcommand{\xmark}{\textcolor{red}{\ding{55}}}%

\begin{document}
\title{MobileNVC: Real-time 1080p Neural Video Compression on a Mobile Device}

\author{
 Ties van Rozendaal, Tushar Singhal, Hoang Le, Guillaume Sautiere, Amir Said, \\
 Krishna Buska, Anjuman Raha,  Dimitris Kalatzis, Hitarth Mehta, Frank Mayer, Liang Zhang, \\ 
 Markus Nagel, Auke Wiggers \\
 {\tt\footnotesize  
   ties@qti.qualcomm.com, auke@qti.qualcomm.com
 } \\
 Qualcomm AI Research\thanks{This text will not show up in this template.}
}

\newcommand{\new}[1]{{\color{red} #1}}

\twocolumn[{
\maketitle
\begin{center}
    \captionsetup{type=figure}
    \includegraphics[width=\linewidth]{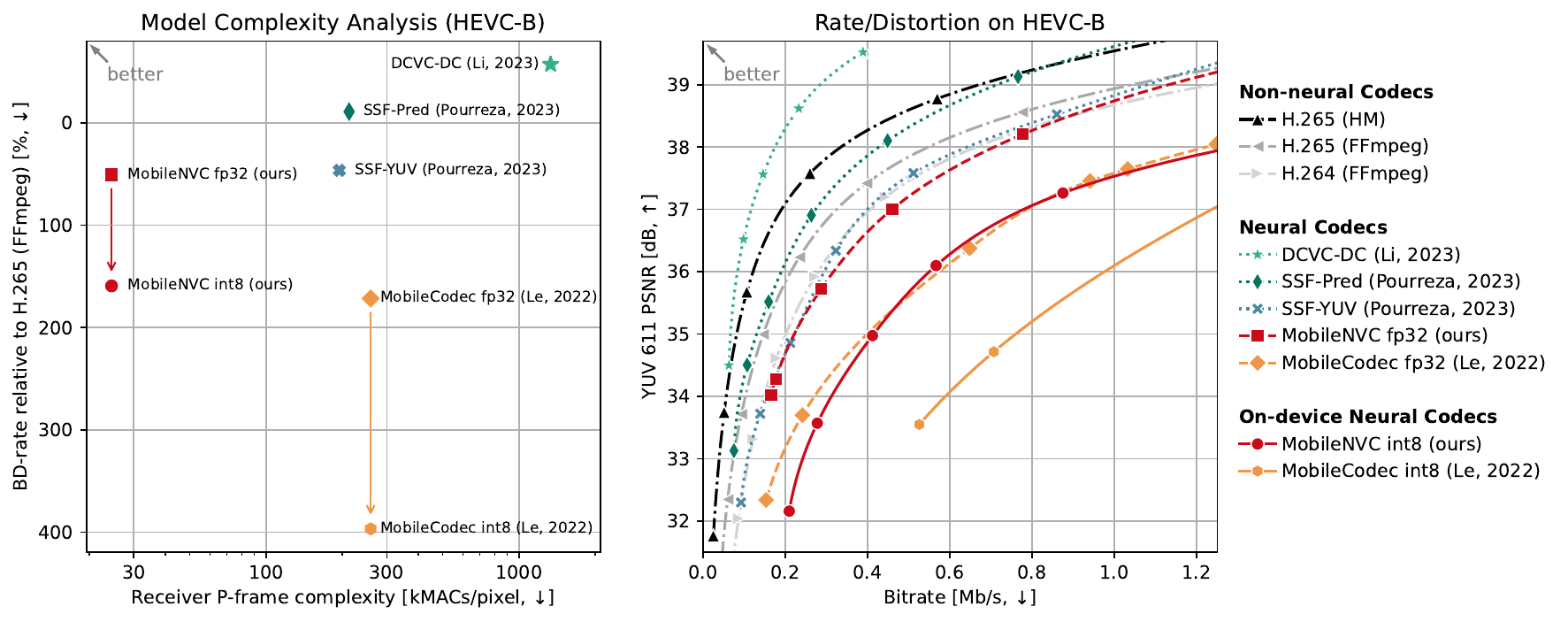}
        \captionof{figure}{Compression performance versus receiver compute (left) and corresponding rate-distortion curves (right). Bjøntegaard Delta-rate is calculated with H.265 (FFmpeg, preset fast) as reference. The kMACS/pixel are computed using a full-HD $ 1080 \times 1920 $ input.}
    \label{fig:teaser}
\end{center}
}]

\renewcommand{\thefootnote}{\fnsymbol{footnote}}
\footnotetext[1]{Qualcomm AI Research is an initiative of Qualcomm Technologies, Inc.} 
\renewcommand*{\thefootnote}{\arabic{footnote}}

\begin{abstract}
    Neural video codecs have recently become competitive with standard codecs such as HEVC in the low-delay setting.
However, most neural codecs are large floating-point networks that use pixel-dense warping operations for temporal modeling, making them too computationally expensive for deployment on mobile devices. 
Recent work has demonstrated that running a neural decoder in real time on mobile is feasible, but shows this only for 720p RGB video.

This work presents the first neural video codec that decodes 1080p YUV420 video in real time on a mobile device.
Our codec relies on two major contributions.
First, we design an efficient codec that uses a block-based motion compensation algorithm available on the warping core of the mobile accelerator, and we show how to quantize this model to integer precision.
Second, we implement a fast decoder pipeline that concurrently runs neural network components on the neural signal processor, parallel entropy coding on the mobile GPU, and warping on the warping core.
Our codec outperforms the previous on-device codec by a large margin with up to 48~\% BD-rate savings, while reducing the MAC count on the receiver side by $10 \times$. 
We perform a careful ablation to demonstrate the effect of the introduced motion compensation scheme, and ablate the effect of model quantization.

\end{abstract}

\section{Introduction}

Neural video compression has seen significant progress in recent years. 
In particular, in the \emph{low-delay P} setting, various works \cite{pourreza2022boosting,li2022hybrid,li2023diversecontexts} have outperformed reference implementations of standard codecs like HEVC (HM) \cite{sullivan2012hevc} and VVC (VTM) \cite{vtm2022} in compression performance.
However, current neural codecs are computationally expensive compared to standard solutions, and reported runtimes are often measured on powerful desktop or datacenter GPUs. 
Additionally, many works assume the availability of pixel-based \cite{agustsson2020ssf, rippel2021elfvc,pourreza2022boosting} or feature-based \cite{hu2021fvc,li2021deep,li2022hybrid,li2023diversecontexts} warping operations, which may be hard to efficiently implement on resource-constrained devices such as mobile phones.

Standard codecs, on the other hand, have fast software implementations such as FFmpeg \cite{wiegand2003h264, sullivan2012hevc}, or efficient silicon implementations specifically designed for fast decoding on consumer hardware.
Although some works design neural codecs with efficiency in mind \cite{rippel2021elfvc,vanrozendaal2023instanceadaptive,shi2022alphavc,galpin2023entropy,mentzer2023m2t}, the only published work that reports runtime on a mobile device is MobileCodec~\cite{le2022mobilecodec}.
One of the main contributions of MobileCodec is to replace optical flow warping with a convolutional motion compensation network, avoiding the need to implement the warping operation on-device.
However, this design has a negative impact on compression performance.

In this work, we build MobileNVC, a neural P-frame codec architecture designed for deployment to a mobile device.
Instead of replacing warping, MobileNVC introduces a block-based warping scheme that can be implemented efficiently using a motion compensation kernel available on the \snapdragon\footnote{Snapdragon branded products are products of Qualcomm Technologies, Inc. and/or its subsidiaries.} 8 Gen 2 neural accelerator. 
Our network design is based on the model architecture of \cite{pourreza2022boosting}, 
made more efficient by using a lean flow extrapolator that predicts the next flow for blocks of pixels, and by using only few warping operations.

We improve inference efficiency by quantizing weights and activations to 8-bit integers.
We show that naive quantization of the mean parameter of a mean-scale hyperprior leads to catastrophic loss in R-D performance, and propose a solution for low-precision quantization.
For the scale parameter, we use the efficient quantization scheme of Said et al. \cite{said2022optimized}.
We further increase throughput by implementing a parallel entropy coding algorithm on GPU, massively increasing parallelism to hundreds of threads \cite{said2023minohead}, compared to the eight threads used by MobileCodec.

Together, these techniques enable extremely efficient neural video decoding on a mobile device.
We obtain 48\% Bj{\o}ntegaard Delta-rate savings compared to MobileCodec, and a $10\times$ reduction in computational complexity.
Additionally, where MobileCodec only decodes HD (720p) video, we enable running $>$30fps full-HD (1080p) real-time decoding on mobile.
We study the effect of the choice of warping operator and quantization in careful ablation studies, allowing us to determine key factors for effective mobile-friendly design of neural video codecs. 
Key results on compression performance and computational efficiency are shown in \Cref{fig:teaser}.

\section{Related work}

\subsection{Neural data compression}

Neural codecs are systems that learn to compress data from examples. 
The most widely adopted model for neural data compression is the mean-scale hyperprior \cite{ballé2018variational, minnen2018joint}.
This model is a hierarchical variational autoencoder with quantized latent variables, and can be seen as  a specific version of a \emph{compressive autoencoder} \cite{theis2017lossy}.

After initial success in the image domain \cite{balle2018variational, rippel2017realtime, minnen2018joint}, neural codecs were extended to the video setting \cite{lu2019dvc, habibian2019video, golinski2020feedback}.
Inspired by standard codecs, neural video codecs adopted motion compensation and residual coding using task-specific networks \cite{lu2019dvc, rippel2019lvc, agustsson2020ssf}.
These architectures were further augmented using predictive models that predict the flow, residual or both \cite{rippel2021elfvc, pourreza2022boosting, hu2022coarse}, leading to improvements in compression performance. 
Recent works show that conditional coding can be more powerful than residual coding \cite{li2022hybrid, li2023diversecontexts}, and perform similarly to the strongest standard video codecs.
However, these architectures require multi-stage training to prevent aggregating error, and introduce various custom operations, such as feature-space motion compensation \cite{hu2021fvc}.

\subsection{Efficient neural video codecs}

Neural codecs are quickly closing the gap with standard codecs, but improved compression performance typically comes with an increase in computational cost \cite{Zhu_Yang_Cohen_2022}.
For this reason, many works now report runtime and show the number of Multiply-Accumulate (MAC) operations.
However, runtime is typically measured on desktop or datacenter GPUs.
The deployment of neural codecs to resource-constrained devices has received relatively little attention.

In the learned image compression setting, early works improved rate-distortion performance by introducing bigger and better prior models \cite{minnen2018joint, wu2020gan, cheng2020learned, guo2021soft}.
Follow-up work reduced computational cost by careful prior model design \cite{he2022elic}, or via transformer-based architectures, where much of the efficiency comes from the ability to parallelize computation across independent sub-tensors \cite{Zhu_Yang_Cohen_2022, mentzer2023m2t}.
Additionally, both Galpin et al. \cite{galpin2023entropy} and Yang et al. \cite{yang2023asymmetrically} show that highly asymmetric encoder-decoder architectures allow using a receiver with much lower MAC count, and 
EVC \cite{wang2023evc} shows that distillation and pruning can also prove effective. 

In the learned video compression setting, various works aim to reduce the computational cost of the receiver
\cite{rippel2021elfvc,vanrozendaal2023instanceadaptive,le2022mobilecodec,shi2022alphavc}. 
For instance, ELF-VC~\cite{rippel2021elfvc} designs a specific convolutional block to improve inference speed and BD-rate. 
AlphaVC introduces a technique that allows skipping tokens during entropy decoding, reducing runtime \cite{shi2022alphavc}.
Van Rozendaal et al.~\cite{rozendaal2021instance, vanrozendaal2023instanceadaptive} show that by overfitting the codec to the instance to compress, one can drastically reduce computational cost on the receiver side.

Nevertheless, most neural video codecs include operations that are difficult to implement efficiently on-devices where the size of the memory is constrained.
Examples include advanced motion compensation algorithms such as \emph{scale-space warping} \cite{agustsson2020ssf,rippel2021elfvc,pourreza2022boosting} or warping in feature-space using deformable convolutions \cite{hu2021fvc, li2022hybrid,  hu2022coarse, li2023diversecontexts, qi2023motion}.
To avoid these operations, some codecs replace motion compensation entirely, 
for example by only modeling the relation between frames via the prior model \cite{mentzer2022vct}.

The main baseline for our work, MobileCodec \cite{le2022mobilecodec}, 
is the only work that demonstrates decoding of video on a mobile device in real-time.
It achieves this by replacing the warping operation by a learned motion compensation sub-network, quantizing the weights and activations, and by implementing parallel entropy coding on mobile CPU
\cite{said2015compressed, said2022optimized}.

\subsection{Quantizing neural codecs}

A common methodology for computational cost reduction is quantization of weights and activations.
For neural \emph{image} compression, one of the first works \cite{balle2018integer} studying neural quantization was mainly motivated by cross-platform reproducibility, as entropy coding is sensitive and may break due to non-deterministic floating-point operations. 
investigated Post-Training Quantization (PTQ) \cite{he2022posttraining,koyuncu2022device,sun2022qlic,shi2022rate} and Quantization-Aware Training (QAT) \cite{sun2020endtoend,hong2021efficient,sun2021endtoend,sun2021learned} techniques for both weights and activations, with the aim to close the rate-distortion gap between the integer-quantized models and their floating-point counterparts.
For instance, Sun et al.~\cite{sun2022qlic} introduce channel splitting, where the convolution output channels most sensitive to quantization are split up and quantized using a custom dynamic range, while other channels are pruned away.
Various works \cite{sun2020endtoend,sun2021learned,sun2022qlic,shi2022rate} have shown that using \emph{per-channel} activation quantization can be effective.
However, this requires bit shifts on the accumulator which is not commonly supported on most fixed-point accelerators.
We therefore use the commonly supported, but less flexible, \emph{per-tensor} quantization for the activations \cite{nagel2021white}.
Note that all works above perform simulated quantization. 
Instead, we implement and benchmark the performance of the quantized model on a mobile device.

\section{Method}

We first describe the architecture, block-based warping scheme, and training losses needed to train a 32-bit floating point model. 
We then describe the quantization procedure, and how we run entropy coding and inference on-device.

\subsection{Network architecture}

The MobileNVC architecture is a variation of the scale-space flow \cite{agustsson2020ssf} architecture of Agustsson et al.
Its input and output heads are modified for YUV 4:2:0 inputs following Pourreza et al. \cite{pourreza2022boosting}.
The motivation for operating in the YUV color space is that distance in this space is better aligned with human perception \cite{tasic2003colorspaces},
and the 4:2:0 subsampling scheme exploits the difference in sensitivity of the human eye between luminance and color. 

\begin{figure}
    \centering

    \includegraphics[width=0.73\linewidth]{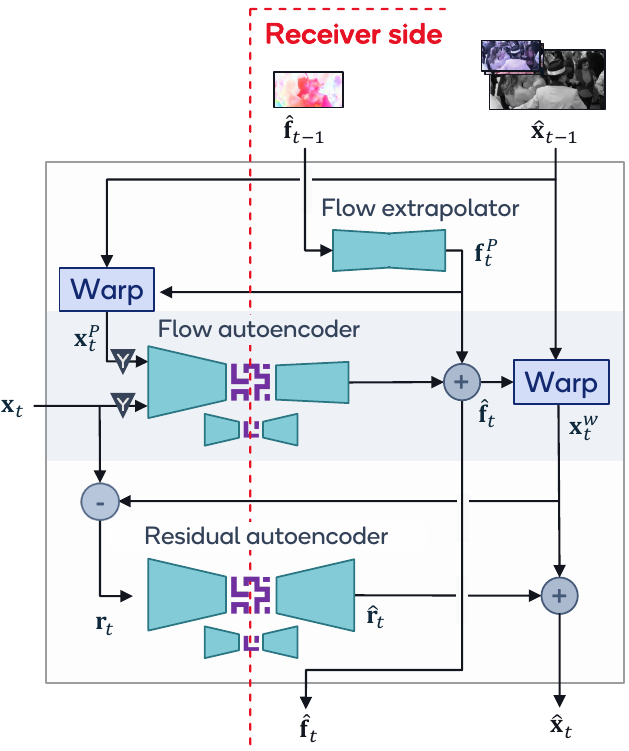}
    \caption{
        Architecture of our P-frame model. Due to the overlap block-based warping, the motion vectors are lower dimensional leading to reduced compute. The flow encoder is further optimized by only using the Y-channel of the inputs. \protect\footnotemark
    }
    \label{fig:flowchart}
\end{figure}
\footnotetext{Image data from Tango video from
Netflix Tango in Netflix El Fuente. Video produced by Netflix, with CC BY-NC-ND 4.0 license: \url{https://media.xiph.org/video/derf/ElFuente/Netflix_Tango_Copyright.txt}}

The model consists of three mean-scale hyperprior \cite{minnen2018joint} autoencoders and a flow extrapolator.
The first mean-scale hyperprior acts as the \emph{I-frame} model, compressing the first frame in a Group of Pictures (GoP) independent of other frames.
The model for each subsequent \emph{P-frame} consists of the remaining two autoencoders and the flow extrapolator, as visualized in \Cref{fig:flowchart}.
Compressing a P-frame consists of three steps.
First, the flow extrapolator predicts a flow $\fvec^P_t$ based on the previously transmitted flow $\hat{\fvec}_{t-1}$ (which we set to zero for frame $x_1$). 
Using this flow, we use warping (on the sender-side) to obtain an initial prediction for the next frame, $\xvec^P_t$.
Second, we transmit a flow delta.
The Y-channels of both $\xvec^P_t$ and the current ground truth frame $\xvec_t$ are given to the flow autoencoder, 
which estimates and compresses the flow residual $\fvec^P_t - \hat{\fvec}_{t}$.
The reconstruction by the flow autoencoder is then added to the extrapolated flow to obtain the refined reconstructed flow $\hat{\fvec}_t$.
We warp the previous predicted frame $\hat{\xvec}_{t-1}$ with $\hat{\fvec}_t$ to form the refined prediction $\xvec^w_t$.
Third, the residual autoencoder compresses the frame residual $\rvec = \xvec_t - \xvec^w_t$.
The resulting reconstruction $\hat{\rvec}$ is added to the warped frame on the receiver-side in order to form the final predicted frame $\hat{\xvec}_t$. 
Full details on network architecture can be found in \Cref{fig:full_model_architecture} in the Appendix.

\subsection{Efficient block-based warping}

Motion compensation is an essential component of both standard and neural video codecs. 
In this work, we use a block-based motion compensation scheme that has two main advantages over pixel-based schemes. 
First, it is possible to implement this scheme efficiently on the mobile neural accelerator.
Second, by warping block-by-block, the flow tensor is of lower spatial dimensionality than the frame, reducing the computational cost of the flow autoencoder and extrapolator networks.

We first describe traditional, pixel-dense optical flow warping.
A frame $\xvec$ is warped using a flow field $\fvec$, which is a 2D map indicating horizontal and vertical displacements.
Specifically, for every pixel $i, j$ in the warped frame, the value is retrieved from the reference frame as follows:
\vspace{-1mm}
\begin{align}
    \DenseWarp{\xvec, \fvec}_{i,j} = \xvec[i + \fvec_x[i,j], j + \fvec_y[i,j]].
\end{align}

Here $[\cdot]$ refers to array indexing, $x$ and $y$ sub-indices indicate retrieval of the respective coordinate in the vector field $\fvec$. 
For non-integer motion vectors, bilinear or bicubic interpolation is typically used to compute the pixel intensity.

The motion vector $\fvec$ often contains large homogeneous regions, as large objects and the background rarely show chaotic motion.
Therefore, block-based warping can be used as a computationally efficient alternative to pixel-space warping.
Here, the warped frame is divided into blocks of size $b \times b$, and all pixels in a block are retrieved from the reference frame using a single shared motion vector. 
The frame is thus warped as follows:
\vspace{-1mm}
\begin{align}
    & \BlockWarp{\xvec, \fvec, b}_{i,j} = \nonumber \\
    & \quad\quad\ 
         \xvec \lb 
              i + \fvec_x\lb \floor{\dfrac{i}{b}}, \floor{\dfrac{j}{b}} \rb, 
              j + \fvec_y \lb \floor{\dfrac{i}{b}}, \floor{\dfrac{j}{b}} \rb 
         \rb.
\end{align}


Block-based warping can be more efficient than dense pixel warping due to the block-wise memory access. 
However, one downside is that artifacts might occur around the block edges when adjacent blocks have different motion vectors. 
This can be solved using \emph{overlapped block motion compensation}.
Here, each block is warped multiple times using the $N-1$ surrounding motion vectors, and the results are averaged using a kernel 
$\mathbf{w} \in \Real^{b \times b \times N}$ 
that decays towards the end of the blocks \cite{nogaki1992overlapped, orchard1994overlapped}, here a Gaussian:

\vspace{-1mm}
\begin{align}
    & \OverlapBlockWarp{\xvec, \fvec, \mathbf{w}, b}_{i,j} = \sum_{k=1}^{N} w_{i,j,k} \cdot  
    \, \xvec \Bigg [
        \nonumber \\
       &\quad i + \fvec_x \lb \floor{\dfrac{i}{b}} + b \cdot \Delta_x^k, \floor{\dfrac{j}{b}} + b \cdot \Delta_y^k \rb, 
        \nonumber \\
       &\quad j + \fvec_y \lb \floor{\dfrac{i}{b}} + b \cdot \Delta_x^k, \floor{\dfrac{j}{b}} + b \cdot \Delta_y^k \rb 
       \Bigg ],
\end{align}
where $\Delta^k$ defines the relative position of the neighboring block, e.g. $(-1, -1)$ for the top-left block, $(-1, 0)$ for the center-left block.

We deploy an overlapped block motion compensation available in the mobile neural accelerator.
As we will show in Section \ref{sec:results:model_ablation}, this leads to better compression performance than block warping, matches that of dense pixel-space warping, and improves computational efficiency.

\subsection{Loss functions}
\label{sec:method:losses}

Models are trained using a loss consisting of a rate term, a distortion term, and auxiliary losses for the flow components.
Similar to previous work, the rate loss is the sum of negative log-likelihoods of the latents and hyperlatents for the three auto-encoders \cite{agustsson2020ssf}.
Specific to our setup is that we use a zero-centered normal distribution with learned variance as the entropy model for the hyper-latent, instead of the non-parametric hyperprior from Ball\'e et al. \cite{balle2018variational}.
This enables us to use the same entropy coding algorithm for latent and hyper-latent.
We use rounding of latents and hyper-latents at evaluation time, but use a ``mixed'' quantization scheme during training: additive noise quantization when computing the rate loss, and rounding when computing the distortion losses \cite{guo2021soft}.
We reweigh the mean squared error (MSE) distortion losses for the Y:U:V channels with weights 6:1:1, to align with the evaluation metrics \cite{strom2020working, pourreza2022boosting}:
\begin{align}
    D(\xvec, \hat{\xvec}) 
    &=  \tfrac{6}{8} \operatorname{MSE}_Y
      + \tfrac{1}{8} \operatorname{MSE}_U
      + \tfrac{1}{8} \operatorname{MSE}_V.
    \label{eq:yuv_D_loss}
\end{align}

One challenge of training small models at lower bitrates is that frame quality deteriorates over time due to error accumulation. 
To account for this, we use an exponentially modulated P-frame loss that places emphasis on later frames, inspired by schemes that weigh the loss for each frame in the GoP differently \cite{rippel2021elfvc, li2022hybrid}: 
\vspace{-2mm}
\begin{align}
    D_\text{mod}(\xvec, \hat{\xvec}, \tau) 
    &= \frac{T}{{\sum_{i=0}^{T-1} \tau^i}} \sum_{i=0}^{T-1} \tau^i D(\xvec_i, \hat{\xvec}_i) 
\end{align}

Additionally, we halve the value of the rate loss multiplier for I-frames, such that the PSNR value for the chosen operating point of I-frames and P-frames becomes more similar \cite{li2021deep}. 
Lastly, we use auxiliary flow losses during training of our floating point model, to force the network to learn meaningful extrapolated and reconstructed flow fields. 
For both flow outputs $\fvec\in\{\fvec^p,\hat{\fvec}\}$, we set
$D_\text{flow}(\fvec, \hat{\xvec}_{t-1}, \xvec_t) = D\lp \warp{\hat{\xvec}_{t-1}, \fvec}, \xvec_t \rp$.
Our final loss is then a weighted combination of all loss terms:

\begin{align}
  \mathcal{L}(\xvec) 
   &= 
     \beta R(\xvec_0) 
     + D(\xvec_0, \hat{\xvec}_0) 
     + 2 \beta R(\xvec_{>0})  
   \nonumber \\ 
     + & D_\text{mod}
     (\xvec_{>0}, \hat{\xvec}_{>0}, \tau) 
     + \lambda D_\text{flow}(\fvec^P) 
     + \lambda D_\text{flow}(\hat{\fvec}).
\label{eq:totallossfunction}
\end{align}

\noindent We train one model for each value of $\beta$.
We show the values of $\lambda$ and $\tau$ for different training stages in \Cref{tab:training_stages} in the Appendix. 

\subsection{Integer Model Quantization}
\label{sec:method:model_quantization}

\begin{figure}
    \centering
    \includegraphics[width=0.99\linewidth]{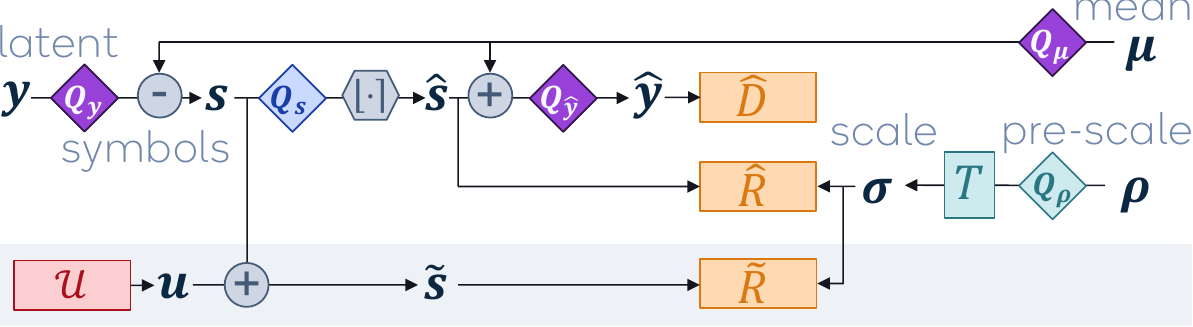}
    \caption{Computational graph of the latent bottleneck and entropy model during 
    training. 
    During floating point training, only the symbols $\svec$ are quantized using rounding (hexagonal rounding operator),
    and a proxy rate loss $\tilde{R}$ based on additive quantization noise $\uvec$ is used (bottom pathway). 
    During quantization-aware training, quantizers (shown as diamonds) are added to the graph. 
    Quantizers with the same color and symbol have tied grids.
    }
    \label{fig:quantization_pipeline}
    \vspace{-2mm}
\end{figure}  

After training a model in 32-bit floating point, we quantize weights and activations to 8-bit integer precision using the AIMET library \cite{siddegowda2022aimet} in two stages.
In the Post-Training Quantization (PTQ) stage, we estimate the quantizer parameters by passing a small amount of data through the network using a per-layer MSE loss~\cite{nagel2021white}.
To improve performance, we then add a Quantization-Aware Training (QAT) stage, where we use LSQ~\cite{lsq} to finetune both the network and quantization parameters using gradient descent. 
The exact hyperparameters of each stage can be found in \Cref{tab:training_stages}. 

We use integer quantization with a learned uniform grid, defined by a step size and a zero offset parameter. 
For the network weights, we learn a grid per output channel without zero offset, i.e., symmetric per-channel quantization. 
For activations, we learn a single quantization grid with a scale and zero offset, i.e., asymmetric \emph{per-tensor} quantization. 

A few operations require custom quantization grids.
As noted in recent works \cite{koyuncu2022device,shi2022rate}, bottleneck quantization in a mean-scale hyperprior architecture needs careful consideration when quantizing activations and performing entropy coding. 
The computational graph of the entropy model for the bottleneck of a mean-scale hyperprior is shown in \Cref{fig:quantization_pipeline}. 
We will first describe this graph without considering model quantization and then describe how we set the quantizers (shown as diamonds in this plot). 

The latents $\yvec$ are the output of the mean-scale hyperprior encoder.
These are not transmitted directly, rather, we transmit the \emph{symbols} $\svec = \yvec - \muvec$, which are the latents after mean subtraction.
During inference, symbols are rounded to the integer grid $\hat{\svec} = \round{\svec}$.
During training, rounding is simulated in a differentiable manner using uniformly sampled additive noise $\uvec \sim U[-0.5, 0.5]$, resulting in $\tilde{\svec} = \svec + \uvec$ (\Cref{fig:quantization_pipeline}, bottom path in grey).
A proxy rate loss $\tilde{R}$ is then based on these noisy latents, while the distortion loss $\hat{D}$ is based on the quantized latents \cite{guo2021soft}.

To quantize the model for on device inference, we add quantizers for each of the variables, indicated by diamonds in \Cref{fig:quantization_pipeline}. 
The work of Said et al. \cite{said2022optimized} shows how to best quantize the scale $\sigmavec$: 
let the network predict a \emph{pre-scale} $\rhovec \in (0, 1]$, which is mapped to the scale using an exponential-polynomial function $\sigmavec = T(\rhovec)$ by the entropy decoder.  
As the domain of this function is fixed, we can fix the quantizer $Q_\rho$ to the same grid.
At inference, $\rhovec$ is directly passed to the entropy coding algorithm in int8.

The remaining question is how to choose quantizers for the latents, mean and symbols. 
As symbols are rounded to the integer grid, we choose their quantizer $Q_s$ to have a symmetric grid with step size 1. 
For the pre-quantized latents $\yvec$ and the mean $\muvec$, we show experimentally that sub-integer precision is required for good rate-distortion performance, both in the PTQ and QAT setting. 
Specifically, we show that this problem can be solved by either using an 8-bit quantizer with a step size $\frac{1}{5}$ ($\frac{1}{3}$ for the highest four bitrates) or by using a 16-bit quantizer for $\yvec, \hat{\yvec}$, and $\muvec$. We observed that careful alignment of the grids performs better than learning a quantization grid via backpropagation.
We emphasize that during the QAT stage, the rate loss is based on latents perturbed with uniform quantization noise.

\begin{figure*}
    \centering
    \includegraphics[width=\linewidth]{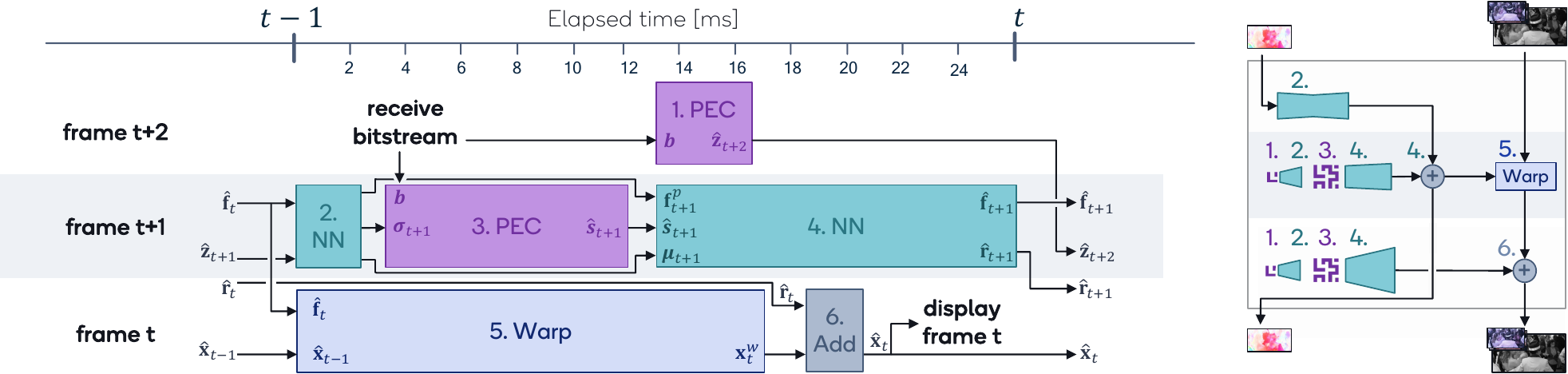}
    \caption{Pipelining of decoding stages. 
        We show each stage and its inputs and outputs. 
        Colors indicate on which component of the chip the stage runs:
        Neural networks (NN) on the neural accelerator, 
        parallel entropy decoding (PEC) on the GPU, 
        overlap block warping (Warp) on the warping kernel, 
        and addition (ADD) on the CPU. 
        Approximate runtime is indicated by the block width.
    }
    \label{fig:pipeline}
\end{figure*}

\subsection{Entropy Coding and Pipelined Inference}

Our decoder must reach a throughput of $30+$ frames per second (FPS) for full HD (1080 x 1920 pixels) YUV 4:2:0 videos. 
To best utilize available compute, we design an inference pipeline that uses different subsystems of the 
mobile neural processing unit (NPU).
This pipeline is shown in \Cref{fig:pipeline}. 
Data from up to three timesteps are processed in parallel by simultaneously using the GPU, NPU, CPU and the warping kernel.
Additionally, where MobileCodec used the CPU to perform entropy coding \cite{le2022mobilecodec}, our GPU implementation can easily be set to run entropy coding exclusively, ensuring stable performance and framerate.

As the mobile NPU and GPU can share memory, there are no delays caused by copying data between processing elements. The amount of entropy coding data to be processed is minimized by using only 8-bit integers for data elements.
Arithmetic coding functions are implemented with OpenCL, and thanks to the small size of their tables, parallelization is defined mostly by the number of OpenCL work items, and work group organization is not critical~\cite{kaeli2015ocl}.

\section{Experiments}

\paragraph{Training stages}
Training consists of two floating point training stages and two quantization stages.
The hyperparameters for these training stages can be found in \Cref{tab:training_stages} in the Appendix. 
The first two stages train the floating-point model using the loss of Equation \ref{eq:totallossfunction}.
In stage one, auxiliary losses for the flow are used, and stage two removes these by setting the loss factor $\lambda=0$.
In stage three, we perform PTQ by fitting the quantizers while keeping the model parameters fixed. 
In stage four, we perform QAT, by finetuning the model and quantizer parameters using LSQ~\cite{lsq}.

\paragraph{Datasets}
We train all models on Vimeo90k~\cite{vimeo90k}.
We use the Xiph-5N dataset \cite{xiph, van2021overfitting} as validation set, and tune hyperparameters on this set.
We evaluate on multiple video compression benchmarks: HEVC-B test sequences~\cite{hevc}, UVG-1k~\cite{uvg} sequences, and MCL-JVC \cite{mcl_jvc} sequences.

\paragraph{Neural Baselines}
We compare our models against state-of-the-art neural video compression methods that report YUV performance like SSF-YUV and SSF-Pred from Pourreza et al. \cite{pourreza2022boosting} and DCVC-DC from Li et al.\cite{li2023diversecontexts}. 
The latter does not report performance on full video sequences, and we re-evaluate their multi-rate model, adjusting the quantization scales for the I-frame and context model to obtain performance at lower bitrates. 

The main neural baseline is MobileCodec \cite{le2022mobilecodec}, as it is the only work to report results from a mobile device implementation. 
This model was not designed for YUV color space. 
We therefore train it on RGB for 1M steps and then finetune it on YUV 6:1:1 R-D loss for another 500k steps.
We use the best available group of pictures (GoP) size for every model. This is GoP=32 for Li et al., GoP=$\infty$ for Pourreza et al., and GoP=16 for our model and MobileCodec.

\paragraph{Standard Baselines} 
We run H.265 and H.264 using FFmpeg \cite{tomar2006converting} with the `fast' preset.
We also run HM \cite{sullivan2012hevc} with default configurations. 
Note that on-device implementations of this standard codec are typically less performant than this reference implementation.
We disable B-frames and use a GoP=$\infty$. 
We do not include VVC or AV1.
Details and exact commands can be found in Appendix \ref{app:standard_codecs_commands}.

\paragraph{Metrics}
To evaluate compression performance, we compute peak signal-to-noise ratio (PSNR) on the Y, U and V channels separately. 
In line with common evaluation protocols \cite{strom2020working} we average PSNR over the Y:U:V channels with weights of 6:1:1. We use Bjøntegaard-Delta bitrate \cite{bjontegaard2001bdrate} (BD-rate) to summarize rate-distorion performance in a single metric, based on bits-per-pixel (bpp) and YUV 6:1:1 PSNR. 

We discard rate-distortion points where the bitrate is unreasonably high (above 1.3 Mb/s).
To ensure a fair comparison, we omit outlier points: we keep points only if all other methods have a point that falls in the same PSNR range.

We compute the MAC count for our models and neural baselines using DeepSpeed \cite{aminabadi2022deepspeed}. Details can be found in Table~\ref{tab:model_complexity_per_subnetwork} in the Appendix.
For the MobileCodec-int8 model \cite{le2022mobilecodec} we report AIMET \cite{siddegowda2022aimet} simulated rate-distortion results.
To evaluate MobileNVC, we do not rely on simulated performance, but rather evaluate the actual rate-distortion resulting from on-device encoding and decoding.

\section{Results and ablations}

\subsection{Rate-distortion and model complexity}
\label{sec:results:compression_performance}

\Cref{fig:teaser} shows BD-rate versus receiver complexity (left), as well as the rate-distortion performance curves (right).
We summarize BD-rate results in \Cref{tab:BD_rates_small_hevc_b}.
The best-performing floating-point neural codec, DCVC-DC, also has the largest model complexity (about $50\times$ more MAC operations than our MobileNVC model). 
Our floating-point model (MobileNVC fp32) has the lowest receiver-side complexity at 24.5 kMACs/pixel, and matches the BD-rate of the SSF-YUV model, despite having an $\sim8$ times lower MAC count.
Our floating point codec still underperforms H.265 (FFmpeg).
However, compared to MobileCodec, which was also designed for on-device inference, we improve compression performance by 45~\% whilst reducing model complexity by more than $10\times$.

\begin{table}[t]
    \footnotesize
    \centering
    \begin{tabular}{llrrrr}
\toprule
& \textbf{BD rate metric}                       & \multicolumn{2}{c}{\textbf{YUV:611 PSNR}}    & \multicolumn{2}{c}{\textbf{Y PSNR}}         
\\
& \textbf{Dataset}                              & \textbf{HEVC-B}      &\textbf{MCL}          & \textbf{HEVC-B}      &\textbf{MCL}         \\
 \midrule
\parbox[t]{0.7mm}{\multirow{2}{*}{\rotatebox[origin=c]{90}{\textbf{int8}}}} 
& MobileNVC                            & \textbf{159.2}    & \textbf{192.6}     & \textbf{124.9}   & \textbf{165.2}    \\
& MobileCodec \cite{le2022mobilecodec} & 396.7              & 549.0               & 319.6              & 570.7              \\
 \midrule
 \parbox[t]{0.7mm}{\multirow{5}{*}{\rotatebox[origin=c]{90}{\textbf{float32}}}} 
& MobileNVC                            & 50.8               & 56.4                & 32.5               & 41.9               \\
& MobileCodec \cite{le2022mobilecodec} & 171.6              & 294.4               & 145.2             & 268.8              \\
& SSF-YUV \cite{pourreza2022boosting}  & 46.4               & 44.2                & 25.7               & 27.2               \\
& SSF-Pred \cite{pourreza2022boosting} & -10.9              & -0.1                & -24.1              & -10.3               \\
& DCVC-DC \cite{li2023diversecontexts} & \textbf{-57.1}     & -                   & -                  & -                  \\
\bottomrule                       
\end{tabular}
\caption{BD-rate saving (in \%) relative to ffmpeg x.265 for HEVC-B and MCL-JCV datasets, lower is better. DCVC-DC is excluded for datasets where PSNR did not overlap with other methods. More benchmarks can be found in \Cref{fig:RD_results_uvg_mcl} in the Appendix.}
    \label{tab:BD_rates_small_hevc_b}
    \vspace{-5mm}
\end{table}

Our quantized codec (MobileNVC int8) far outperforms the quantized MobileCodec int8 -- the only other work to show real-time mobile video decoding -- with 48~\% BD-rate savings. 
The performance gap between quantized models and state-of-the-art floating point codecs is substantial, but this is not surprising, as being able to decode on a mobile device poses tight computational constraints. 


As we optimized for receiver inference speed, our receiver has 25\% of the kMACs of our sender, whereas for other models this is 70-80\%. 
For our model, components dealing with motion are lower complexity than baselines due to the low-dimensional flow vectors, and the fact that motion-autoencoder only uses Y-channels. 
Warping operations are not taken into account in the MAC count, but as we show in Figure~\ref{fig:pipeline}, it can be executed in parallel with neural inference, without runtime overhead.
More details on model complexity can be found  
in \Cref{tab:model_complexity_per_subnetwork} in the Appendix.

\subsection{Inference Speed}
\label{sec:results:inference_speed}

We measure inference speed on a mobile phone with a Snapdragon 8 Gen 2 system-on-chip.
On HEVC-B, we achieve an average receiver inference speed of $38.9$ FPS.
As our focus is receiver-side complexity, we did not optimize our transmitter pipeline and run all steps sequentially, resulting in an encoding rate of around 3 FPS.
For decoding, \Cref{fig:pipeline} shows the approximate duration of each step. 
Inference speed is bounded by network inference, and due to parallelization, the warping operation is not causing any overhead. 
\Cref{tab:pec_num_threads} shows the speed of our parallel entropy coding implementation. 
The GPU allows us to greatly optimize coding speed by using a large number of threads, at the cost of using more bits due to the larger header. 
We choose to use 512 threads and implement the header naïvely, noting that an optimized implementation could reduce the bitrate overhead by 2x.
All in all, our method not only improves the compression performance compared with MobileCodec but also the throughput, allowing us to operate on full-HD ($1080 \times 1920$) instead of HD ($720 \times 1280$) resolution.

\begin{table}[t]
    \footnotesize
    \centering
    \begin{tabular}{rlrrr}
\toprule
                     &                 &                        & \multicolumn{2}{c}{\textbf{Bitrate overhead } [$\downarrow$]} \\
                     \cmidrule{4-5}
\textbf{\# Threads} & \textbf{Device} & \textbf{Decoding time } [$\downarrow$] & \textbf{Naïve}     & \textbf{Optimized}    \\
\midrule
{1}           & {CPU}    & 20 ms                  & 0.0 \%             & 0.0 \%                \\
{8}           & {CPU}    & 16 ms                  & 0.6 \%             & 0.6 \%                \\
\midrule
{256}         & {GPU}    & 18 ms                   & 2.1 \%             & 1.1 \%                \\
\textbf{512}         & \textbf{GPU}    & \textbf{11 ms}        & \textbf{3.9 \%}    & 2.0 \%       \\
{1024}        & {GPU}    & 6 ms                    & 7.0 \%             & 3.6 \%            \\
\bottomrule 
\end{tabular}
    \caption{Inference speed and rate overhead for different parallelization strategies for entropy coding of the latents. The row in bold indicates the settings we used in our work.}
    \label{tab:pec_num_threads}
\end{table}

\begin{table*}[t]
    \footnotesize
    \centering
    \begin{tabular}{lrllllrrlrrr}
    \toprule
    \textbf{}                              & \textbf{} & \textbf{}               & \multicolumn{2}{l}{\textbf{Model architecture}} & \textbf{} & 
    \multicolumn{2}{l}{\textbf{Params} [M, $\downarrow$]} & \textbf{} & \multicolumn{2}{l}{\textbf{kMACs/px} [$\downarrow$]} & \textbf{}        \\
    \cmidrule{4-5}\cmidrule{7-8}\cmidrule{10-11}
                                           &           & \textbf{Model}          & \textbf{warping}      & \textbf{prior}     & \textbf{} & \textbf{send}     & \textbf{recv}     & \textbf{} & \textbf{send}      & \textbf{recv}      & \textbf{BD-rate} [$\downarrow$] \\
    \midrule                                       
    \textbf{Baseline}      & I.         & MobileNVC, no finetuning                  & overlap block              & mean-scale               &           & 12.42                & 6.30                 &           & 64.93                 & 24.52                   & 0.0~\%                
    \\[6pt]
    \textbf{Warping}
                                           & II.        & block warp              & block warp             & mean-scale               &          & 12.63                 & 6.30                 &           & 64.93                  & 24.52                  & 19.2~\%           \\
                                           & III.       & dense warp              & dense warp            & mean-scale               &          & 12.50                & 6.23                 &           & 153.67                 & 113.59                  & -6.0~\%                \\
                                           & IV.        & conditional conv & no warp               & mean-scale               &           & 9.77                & 6.12                 &           & 80.72                 & 57.35                  & 51.2~\%           
    \\[6pt]
  \textbf{Prior}                         & V.        & scale-only prior        & overlap block              & scale                   &           & 11.86                & 5.74                 &           & 63.84	                 & 23.44                  & 9.6~\%                                                          
    \\[6pt]
    \textbf{Training}                  & VI.       & MobileNVC (+ finetuning)       & overlap block              & mean-scale                   &           & 12.42                 & 6.30                 &           & 64.93                  & 24.52                  & -14.0~\%     \\     
    \bottomrule
\end{tabular}
    \caption{Model architecture ablation. All models are floating-point and have been trained for 1M steps, except for model VI, which trains with stages 1 and 2 (see \Cref{tab:training_stages} for details). 
    Parameters and kMACS/px are shown for the P-frame model only, and are computed for a $1080 \times 1920$ YUV420 input frame. We refer to the Appendix for corresponding R-D-curves (\Cref{fig:RD_model_ablations}, Left).}
    \label{tab:architecture_ablation}
\end{table*}

\subsection{Model Ablations}
\label{sec:results:model_ablation}

We ablate model design choices in \Cref{tab:architecture_ablation}. 
All models in this table are unquantized and are trained only for 1M steps.

First, we look into warping. 
The model with dense warping (row III) has better R-D performance than our overlapped block-warp model (row I). 
However, the gap is small, with 6~\% BD-rate cost, and the dense warp model has more than 4x more MACs due to the higher flow dimensionality. 
Comparing \emph{overlapped} block-warp to \emph{vanilla} block warp (II) without overlap, we see the effectiveness of overlapping, which brings about 19~\% BD-rate savings.
Alternatively, one could use a flow-agnostic model as is done in MobileCodec \cite{le2022mobilecodec}. 
In row IV, we include a variant of our network that uses a conditional convolutional network that can model warping implicitly, as shown in \Cref{fig:model_arch_flow_agnostic} in the Appendix. 
With more than 50~\% BD-rate increase compared to overlapped block-warp, it is suboptimal both in terms of compute and compression performance, showing the importance of warping. An example warped frame for each of the methods can be seen in \cref{tab:warping_samples} in the Appendix.

Next, we look into the probability model for the prior.
We train a version of our model with a scale-only prior (V) instead of a mean-scale prior (I), as in MobileCodec. 
Compression performance is significantly reduced, with a 9.6~\% increase in BD-rate, while efficiency gains are minimal.

Lastly, we quantify the effect of our second training stage, which increases GoP size and uses P-frame loss modulation as described in section \ref{sec:method:losses}. 
Row VI shows that finetuning the main model (row I) for 250K steps with this scheme results in 14~\% BD-rate savings.

\subsection{Quantization ablation}

\label{sec:results:quantization_ablation}

\begin{table*}[t!]
    \footnotesize
    \centering
    \begin{tabular}{rlllllllllllrr}
\toprule
              & \multicolumn{1}{c}{\textbf{Quantization Strategy}}                &  & \multicolumn{8}{c}{\textbf{Quantizer setup}}                                                                     & \textbf{} & \multicolumn{2}{c}{\textbf{BD-rate} [$\downarrow$]} \\[1pt]
              &                                                      &  & \textbf{W} & \textbf{A} & \textbf{} & \raisebox{-2pt}{\includegraphics[height=13pt]{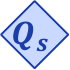}} & \raisebox{-2pt}{\includegraphics[height=13pt]{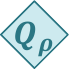}} & \raisebox{-2pt}{\includegraphics[height=13pt]{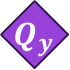}}  & \raisebox{-2pt}{\includegraphics[height=13pt]{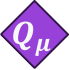}} & \raisebox{-2pt}{\includegraphics[height=13pt]{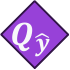}} & \textbf{} & \textbf{PTQ}             & \textbf{QAT}             \\
\cmidrule{1-2}
\cmidrule{4-11}
\cmidrule{13-14}              
{I.}   & Symbols and flow only                       &  & \xmark     & \xmark    &           & \cmark      & \xmark        & \xmark       & \xmark       & \xmark         &           & 11.9~\%                  & -                        \\
{II.}  & + weights                                   &  & \cmark     & \xmark    &           & \cmark      & \xmark        & \xmark       & \xmark       & \xmark         &           & 14.2~\%                  & -                        \\
{III.} & + activations                               &  & \cmark     & \cmark     &           & \cmark      & \xmark        & \xmark       & \xmark       & \xmark         &           & 54.2~\%                  & -                        \\
{IV.}  & + scale                                     &  & \cmark     & \cmark     &           & \cmark      & \cmark        & \xmark       & \xmark       & \xmark         &           & 54.2~\%                  & -                        \\
\midrule
{V.}   & Fully quantized (latent step size=$1$)        &  & \cmark     & \cmark     &           & \cmark      & \cmark        & \multicolumn{3}{c}{\xfill{.6pt} \cmark step size 1~\xfill{.6pt} }       &           & 243.3~\%                 & -                        \\
{VI.}  & Fully quantized (latent step size=$\tfrac{1}{5}$ / $\frac{1}{3}$)     &  & \cmark     & \cmark     &           & \cmark      & \cmark        & \multicolumn{3}{c}{\xfill{.5pt} \cmark step size $\tfrac{1}{5}$ / $\frac{1}{3}$~\xfill{.5pt} }    &           & 101.9~\%                 & 42.5~\%                  \\
{VII.} & Fully quantized (latents int16)             &  & \cmark     & \cmark     &           & \cmark      & \cmark        & \multicolumn{3}{c}{\xfill{.9pt} \cmark int16~\xfill{.9pt} }             &           & 54.2~\%                  & 25.6~\%                  \\
\bottomrule
\end{tabular}
    \caption{Quantization ablation. Values with \xmark \, are unquantized and \cmark indicates values are quantized to int8.
    BD-rate is computed with respect to the floating point baseline on the Xiph-5N dataset (lower is better). R-D curves can be found in \Cref{fig:RD_model_ablations} (Right) in the Appendix.
    }
    \label{tab:quantization_ablation}
\end{table*}

\Cref{fig:teaser} shows that moving from floating point to int8 substantially reduces compression performance.
We break down this reduction in \Cref{tab:quantization_ablation}. 
Row I shows the effect of quantizing symbols $\svec$ and flow vectors $\fvec^P, \hat{\fvec}$.
This leads to a 11.9~\% increase in BD-rate (i.e., worse compression performance), mainly due to flow quantization, and provides an upper bound for quantization performance. 
In row II, we also quantize the weights using per-channel quantization grids, leading to a 14.2~\% overall increase. 
When we also quantize all activations except for those in the latent bottleneck (row III), BD-rate increases to 54.2~\%.

Due to interaction between rounding of the latent symbols (or adding uniform noise) and activation quantization, care should be taken when quantizing the latent bottleneck. 
Row IV shows that the scale-quantization of \cite{said2022optimized} allows us to quantize the scale to 8-bit with no loss in performance.
As described in section \ref{sec:method:model_quantization}, the symbols $\svec$ are rounded to the integer grid, so we use a step size of 1 for their quantization. 
Using the same integer grid for the latents $\yvec$ and the mean $\muvec$ as for the symbols causes a dramatic drop in compression performance (row V), showing the sensitivity of the mean-scale hyperprior to quantization.

One way to overcome this large quantization gap is to use high precision quantization (int16) for the latents and the mean (row VII), but this would increase on-device runtime.
We show in row VI that a carefully chosen quantization grid, with a quantization step size of $\frac{1}{5}$ for the latents and mean parameter, and a step size of $\frac{1}{3}$ for the highest three bitrates, results in a big performance improvement as well. 
This choice allows us to cover a sufficiently large range of values, and avoids that points on the sub-integer grid fall exactly between two points on the coarser grid, thereby reducing ``tie-break'' issues compared to for example using step size $\frac{1}{4}$.
As this choice only requires int8 activations, it does not result in a runtime increase.

Lastly, we see in the rightmost column that compression performance after post-training quantization (PTQ) can be further improved using quantization-aware training (QAT).
Row VI gives us our final BD-rate overhead of 42.5~\% relative to the floating-point model.
Row VII shows that if the runtime increase were acceptable, mixed precision would be the better choice from a compression performance perspective. 
For additional analysis, we refer the reader the rate-distortion curves in \Cref{fig:RD_model_ablations} (right) in the Appendix.

\section{Conclusion}

In this work, we introduce a practical neural codec that performs real-time decoding of full HD video on a mobile device.
This codec (MobileNVC), outperforms the previous state-of-the-art practical codec by 48\% BD-rate, while reducing the Multiply-Accumulate count by $10\times$.
We design an efficient network architecture using a new block-based motion compensation algorithm, and show how to pipeline inference to enable real time decoding on device. Careful ablations show the effect of the introduced motion compensation and quantization schemes. 

Most neural codecs are still too computationally expensive to be adopted in real life settings.
We therefore hope that this work advances the field of practical neural codecs, and that it encourages future authors to benchmark their compression algorithms on-device.

\paragraph{Acknowledgements}
We thank Reza Pourreza for help with training and adapting the SSFPred model.
Thanks to Alireza Shoa, Asma Qureshi and Darren Gnanapragasam for advice on 
the warping kernel.

{\small
\bibliographystyle{wacv_24/ieee_fullname}
\bibliography{references}

\begin{thebibliography}{10}\itemsep=-1pt

\bibitem{agustsson2020ssf}
Eirikur Agustsson, David Minnen, Nick Johnston, Johannes Balle, Sung~Jin Hwang,
  and George Toderici.
\newblock Scale-space flow for end-to-end optimized video compression.
\newblock In {\em Proceedings of the IEEE conference on Computer Vision and
  Pattern Recognition}, 2020.

\bibitem{aminabadi2022deepspeed}
Reza~Yazdani Aminabadi, Samyam Rajbhandari, Minjia Zhang, Ammar~Ahmad Awan,
  Cheng Li, Du Li, Elton Zheng, Jeff Rasley, Shaden Smith, Olatunji Ruwase,
  et~al.
\newblock Deepspeed inference: Enabling efficient inference of transformer
  models at unprecedented scale.
\newblock {\em arXiv preprint arXiv:2207.00032}, 2022.

\bibitem{balle2018integer}
Johannes Ball{\'e}, Nick Johnston, and David Minnen.
\newblock Integer networks for data compression with latent-variable models.
\newblock In {\em International Conference on Learning Representations (ICLR)},
  2018.

\bibitem{balle2018variational}
Johannes Ball{\'e}, David Minnen, Saurabh Singh, Sung~Jin Hwang, and Nick
  Johnston.
\newblock Variational image compression with a scale hyperprior.
\newblock In {\em International Conference on Learning Representations}, 2018.

\bibitem{ballé2018variational}
Johannes Ballé, David Minnen, Saurabh Singh, Sung~Jin Hwang, and Nick
  Johnston.
\newblock Variational image compression with a scale hyperprior.
\newblock In {\em International Conference on Learning Representations}, 2018.

\bibitem{bjontegaard2001bdrate}
Gisle Bj{\o}ntegaard.
\newblock Calculation of average psnr differences between rd-curves.
\newblock 2001.

\bibitem{cheng2020learned}
Zhengxue Cheng, Heming Sun, Masaru Takeuchi, and Jiro Katto.
\newblock Learned image compression with discretized gaussian mixture
  likelihoods and attention modules.
\newblock In {\em Proceedings of the IEEE/CVF Conference on Computer Vision and
  Pattern Recognition}, pages 7939--7948, 2020.

\bibitem{lsq}
Steven~K. Esser, Jeffrey~L. McKinstry, Deepika Bablani, Rathinakumar Appuswamy,
  and Dharmendra~S. Modha.
\newblock Learned step size quantization.
\newblock In {\em International Conference on Learning Representations (ICLR)},
  2020.

\bibitem{galpin2023entropy}
Franck Galpin, M. Balcilar, Fr{\'e}d{\'e}ric Lef{\`e}bvre, Fabien Racap'e, and
  Pierre Hellier.
\newblock Entropy coding improvement for low-complexity compressive
  auto-encoders.
\newblock 2023.

\bibitem{golinski2020feedback}
Adam Golinski, Reza Pourreza, Yang Yang, Guillaume Sautiere, and Taco~S Cohen.
\newblock Feedback recurrent autoencoder for video compression.
\newblock In {\em Proceedings of the Asian Conference on Computer Vision},
  2020.

\bibitem{guo2021soft}
Zongyu Guo, Zhizheng Zhang, Runsen Feng, and Zhibo Chen.
\newblock Soft then hard: Rethinking the quantization in neural image
  compression.
\newblock In {\em International Conference on Machine Learning}, pages
  3920--3929. PMLR, 2021.

\bibitem{habibian2019video}
Amirhossein Habibian, Ties~van Rozendaal, Jakub~M Tomczak, and Taco~S Cohen.
\newblock Video compression with rate-distortion autoencoders.
\newblock In {\em IEEE International Conference on Computer Vision}, 2019.

\bibitem{he2022posttraining}
Dailan He, Ziming Yang, Yuan Chen, Qi Zhang, Hongwei Qin, and Yan Wang.
\newblock Post-training quantization for cross-platform learned image
  compression, 2022.

\bibitem{he2022elic}
Dailan He, Ziming Yang, Weikun Peng, Rui Ma, Hongwei Qin, and Yan Wang.
\newblock Elic: Efficient learned image compression with unevenly grouped
  space-channel contextual adaptive coding.
\newblock In {\em Proceedings of the IEEE/CVF Conference on Computer Vision and
  Pattern Recognition}, pages 5718--5727, 2022.

\bibitem{hevc}
{HEVC}.
\newblock Common test conditions and software reference configurations.
\newblock
  \url{http://phenix.it-sudparis.eu/jct/doc_end_user/current_document.php?id=7281},
  2013.

\bibitem{hong2021efficient}
Weixin Hong, Tong Chen, Ming Lu, Shiliang Pu, and Zhan Ma.
\newblock Efficient neural image decoding via fixed-point inference.
\newblock {\em IEEE Transactions on Circuits and Systems for Video Technology},
  31(9):3618--3630, 2021.

\bibitem{hu2022coarse}
Zhihao Hu, Guo Lu, Jinyang Guo, Shan Liu, Wei Jiang, and Dong Xu.
\newblock Coarse-to-fine deep video coding with hyperprior-guided mode
  prediction.
\newblock In {\em Proceedings of the IEEE/CVF Conference on Computer Vision and
  Pattern Recognition}, pages 5921--5930, 2022.

\bibitem{hu2021fvc}
Zhihao Hu, Guo Lu, and Dong Xu.
\newblock {FVC}: A new framework towards deep video compression in feature
  space.
\newblock In {\em Proceedings of the {IEEE/CVF} Conference on Computer Vision
  and Pattern Recognition}, pages 1502--1511, 2021.

\bibitem{kaeli2015ocl}
David Kaeli, Perhaad Mistry, Dana Schaa, and Dong~Ping Zhang.
\newblock {\em Heterogeneous Computing with {OpenCL~2.0}}.
\newblock Morgan Kaufmann Publishers, Waltham, {MA}, 2015.

\bibitem{koyuncu2022device}
Esin Koyuncu, Timofey Solovyev, Elena Alshina, and André Kaup.
\newblock Device interoperability for learned image compression with weights
  and activations quantization.
\newblock In {\em 2022 Picture Coding Symposium (PCS)}, pages 151--155, 2022.

\bibitem{le2022mobilecodec}
Hoang Le, Liang Zhang, Amir Said, Guillaume Sauti{\`{e}}re, Yang Yang, Pranav
  Shrestha, Fei Yin, Reza Pourreza, and Auke Wiggers.
\newblock Mobilecodec: neural inter-frame video compression on mobile devices.
\newblock In {\em MMSys}, pages 324--330. {ACM}, 2022.

\bibitem{li2021deep}
Jiahao Li, Bin Li, and Yan Lu.
\newblock Deep contextual video compression, 2021.

\bibitem{li2022hybrid}
Jiahao Li, Bin Li, and Yan Lu.
\newblock Hybrid spatial-temporal entropy modelling for neural video
  compression.
\newblock In {\em Proceedings of the 30th ACM International Conference on
  Multimedia}, pages 1503--1511, 2022.

\bibitem{li2023diversecontexts}
Jiahao Li, Bin Li, and Yan Lu.
\newblock Neural video compression with diverse contexts.
\newblock {\em arXiv preprint arXiv:2302.14402}, 2023.

\bibitem{lu2019dvc}
Guo Lu, Wanli Ouyang, Dong Xu, Xiaoyun Zhang, Chunlei Cai, and Zhiyong Gao.
\newblock {DVC}: An end-to-end deep video compression framework.
\newblock In {\em Proceedings of the IEEE conference on Computer Vision and
  Pattern Recognition}, 2019.

\bibitem{mentzer2023m2t}
Fabian Mentzer, Eirikur Agustsson, and Michael Tschannen.
\newblock M2t: Masking transformers twice for faster decoding, 2023.

\bibitem{mentzer2022vct}
Fabian Mentzer, George Toderici, David Minnen, Sergi Caelles, Sung~Jin Hwang,
  Mario Lucic, and Eirikur Agustsson.
\newblock {VCT}: A video compression transformer.
\newblock In Alice~H. Oh, Alekh Agarwal, Danielle Belgrave, and Kyunghyun Cho,
  editors, {\em Advances in Neural Information Processing Systems}, 2022.

\bibitem{uvg}
Alexandre Mercat, Marko Viitanen, and Jarno Vanne.
\newblock {UVG} dataset: 50/120fps 4k sequences for video codec analysis and
  development.
\newblock In {\em ACM Multimedia Systems Conference}, pages 297--302, 2020.

\bibitem{minnen2018joint}
David Minnen, Johannes Ball{\'e}, and George~D Toderici.
\newblock Joint autoregressive and hierarchical priors for learned image
  compression.
\newblock {\em Advances in neural information processing systems}, 31, 2018.

\bibitem{nagel2021white}
Markus Nagel, Marios Fournarakis, Rana~Ali Amjad, Yelysei Bondarenko, Mart van
  Baalen, and Tijmen Blankevoort.
\newblock A white paper on neural network quantization.
\newblock {\em arXiv preprint arXiv:2106.08295}, 2021.

\bibitem{nogaki1992overlapped}
Satoshi Nogaki and Mutsumi Ohta.
\newblock An overlapped block motion compensation for high quality motion
  picture coding.
\newblock In {\em [Proceedings] 1992 IEEE International Symposium on Circuits
  and Systems}, volume~1, pages 184--187. IEEE, 1992.

\bibitem{orchard1994overlapped}
Michael~T Orchard and Gary~J Sullivan.
\newblock Overlapped block motion compensation: An estimation-theoretic
  approach.
\newblock {\em IEEE Transactions on Image Processing}, 3(5):693--699, 1994.

\bibitem{pourreza2022boosting}
Reza Pourreza, Hoang Le, Amir Said, Guillaume Sauti{\`{e}}re, and Auke Wiggers.
\newblock Boosting neural video codecs by exploiting hierarchical redundancy.
\newblock {\em CoRR}, abs/2208.04303, 2022.

\bibitem{qi2023motion}
Linfeng Qi, Jiahao Li, Bin Li, Houqiang Li, and Yan Lu.
\newblock Motion information propagation for neural video compression.
\newblock In {\em Proceedings of the IEEE/CVF Conference on Computer Vision and
  Pattern Recognition}, pages 6111--6120, 2023.

\bibitem{rippel2021elfvc}
Oren Rippel, Alexander~G Anderson, Kedar Tatwawadi, Sanjay Nair, Craig Lytle,
  and Lubomir Bourdev.
\newblock {ELF-VC: Efficient Learned Flexible-Rate Video Coding}.
\newblock {\em Neural Information Processing Systems}, 2021.

\bibitem{rippel2017realtime}
Oren Rippel and Lubomir Bourdev.
\newblock {Real-Time} adaptive image compression.
\newblock In Doina Precup and Yee~Whye Teh, editors, {\em Proceedings of the
  34th International Conference on Machine Learning}, volume~70 of {\em
  Proceedings of Machine Learning Research}, pages 2922--2930. PMLR, 2017.

\bibitem{rippel2019lvc}
Oren Rippel, Sanjay Nair, Carissa Lew, Steve Branson, Alexander~G. Anderson,
  and Lubomir Bourdev.
\newblock Learned video compression.
\newblock In {\em IEEE International Conference on Computer Vision}, October
  2019.

\bibitem{said2023minohead}
Amir Said, Hoang Le, and Farzad Farhadzadeh.
\newblock Bitstream organization for parallel entropy coding on neural
  network-based video codecs.
\newblock In {\em IEEE Int. Symp. on Multimedia}, Dec. 2023.

\bibitem{said2015compressed}
Amir Said, Abo-Talib Mahfoodh, and Sehoon Yea.
\newblock Compressed data organization for high throughput parallel entropy
  coding.
\newblock In {\em Applications of Digital Image Processing XXXVIII}, volume
  9599, pages 528--536. SPIE, 2015.

\bibitem{said2022optimized}
Amir Said, Reza Pourreza, and Hoang Le.
\newblock Optimized learned entropy coding parameters for practical
  neural-based image and video compression.
\newblock In {\em 2022 IEEE International Conference on Image Processing
  (ICIP)}, pages 661--665. IEEE, 2022.

\bibitem{shi2022rate}
Junqi Shi, Ming-Tse Lu, and Zhan Ma.
\newblock {Rate-Distortion Optimized Post-Training Quantization for Learned
  Image Compression}.
\newblock 2022.

\bibitem{shi2022alphavc}
Yibo Shi, Yunying Ge, Jing Wang, and Jue Mao.
\newblock {AlphaVC: High-Performance and Efficient Learned Video Compression},
  2022.

\bibitem{siddegowda2022aimet}
Sangeetha Siddegowda, Marios Fournarakis, Markus Nagel, Tijmen Blankevoort,
  Chirag Patel, and Abhijit Khobare.
\newblock Neural network quantization with ai model efficiency toolkit (aimet).
\newblock {\em arXiv preprint arXiv:2201.08442}, 2022.

\bibitem{strom2020working}
J Str{\"o}m, K Andersson, R Sj{\"o}berg, A Segall, F Bossen, G Sullivan, JR
  Ohm, and A Tourapis.
\newblock Working practices using objective metrics for evaluation of video
  coding efficiency experiments.
\newblock {\em ITU-T and ISO/IEC, JTC}, 1:23002--8, 2020.

\bibitem{sullivan2012hevc}
Gary~J. Sullivan, Jens-Rainer Ohm, Woo-Jin Han, and Thomas Wiegand.
\newblock Overview of the high efficiency video coding (hevc) standard.
\newblock {\em IEEE Transactions on Circuits and Systems for Video Technology},
  22(12):1649--1668, 2012.

\bibitem{sun2020endtoend}
Heming Sun, Zhengxue Cheng, Masaru Takeuchi, and Jiro Katto.
\newblock End-to-end learned image compression with fixed point weight
  quantization.
\newblock In {\em 2020 IEEE International Conference on Image Processing
  (ICIP)}, pages 3359--3363, 2020.

\bibitem{sun2021endtoend}
Heming Sun, Lu Yu, and Jiro Katto.
\newblock End-to-end learned image compression with quantized weights and
  activations, 2021.

\bibitem{sun2021learned}
Heming Sun, Lu Yu, and Jiro Katto.
\newblock Learned image compression with fixed-point arithmetic.
\newblock In {\em 2021 Picture Coding Symposium (PCS)}, pages 1--5, 2021.

\bibitem{sun2022qlic}
Heming Sun, Lu Yu, and Jiro Katto.
\newblock Q-lic: Quantizing learned image compression with channel splitting.
\newblock {\em IEEE Transactions on Circuits and Systems for Video Technology},
  page 1–1, 2022.

\bibitem{theis2017lossy}
Lucas Theis, Wenzhe Shi, Andrew Cunningham, and Ferenc Husz{\'a}r.
\newblock {Lossy image compression with compressive autoencoders}.
\newblock {\em International Conference on Learning Representations}, 2017.

\bibitem{tasic2003colorspaces}
M Tkalcic and J~F Tasic.
\newblock Colour spaces: perceptual, historical and applicational background.
\newblock In {\em The {IEEE} Region 8 {EUROCON} 2003. Computer as a Tool.},
  volume~1, pages 304--308 vol.1, Sept. 2003.

\bibitem{tomar2006converting}
Suramya Tomar.
\newblock Converting video formats with ffmpeg.
\newblock {\em Linux Journal}, 2006(146):10, 2006.

\bibitem{rozendaal2021instance}
Ties van Rozendaal, Johann Brehmer, Yunfan Zhang, Reza Pourreza, and Taco~S
  Cohen.
\newblock Instance-adaptive video compression: Improving neural codecs by
  training on the test set.
\newblock {\em arXiv preprint arXiv:2111.10302}, 2021.

\bibitem{vanrozendaal2023instanceadaptive}
Ties van Rozendaal, Johann Brehmer, Yunfan Zhang, Reza Pourreza, Auke~J.
  Wiggers, and Taco Cohen.
\newblock Instance-adaptive video compression: Improving neural codecs by
  training on the test set.
\newblock {\em Transactions on Machine Learning Research}, 2023.
\newblock Expert Certification.

\bibitem{van2021overfitting}
Ties van Rozendaal, Iris~AM Huijben, and Taco~S Cohen.
\newblock Overfitting for fun and profit: Instance-adaptive data compression.
\newblock In {\em International Conference on Learning Representations
  {(ICLR)}}, 2021.

\bibitem{vtm2022}
{VTM}.
\newblock Vtm reference software for versatile video coding (vvc).
\newblock \url{https://vcgit.hhi.fraunhofer.de/jvet/VVCSoftware_VTM/}, 2013.

\bibitem{wang2023evc}
Guo-Hua Wang, Jiahao Li, Bin Li, and Yan Lu.
\newblock Evc: Towards real-time neural image compression with mask decay.
\newblock {\em arXiv preprint arXiv:2302.05071}, 2023.

\bibitem{mcl_jvc}
Haiqiang Wang, Weihao Gan, Sudeng Hu, Joe~Yuchieh Lin, Lina Jin, Longguang
  Song, Ping Wang, Ioannis Katsavounidis, Anne Aaron, and C-C~Jay Kuo.
\newblock Mcl-jcv: a jnd-based h. 264/avc video quality assessment dataset.
\newblock In {\em 2016 IEEE international conference on image processing
  (ICIP)}, pages 1509--1513. IEEE, 2016.

\bibitem{wiegand2003h264}
T. Wiegand, G.J. Sullivan, G. Bjontegaard, and A. Luthra.
\newblock Overview of the h.264/avc video coding standard.
\newblock {\em IEEE Transactions on Circuits and Systems for Video Technology},
  2003.

\bibitem{wu2020gan}
Lirong Wu, Kejie Huang, and Haibin Shen.
\newblock A {GAN}-based tunable image compression system.
\newblock In {\em Proceedings of the IEEE/CVF Winter Conference on Applications
  of Computer Vision}, pages 2334--2342, 2020.

\bibitem{xiph}
{Xiph.org}.
\newblock Video test media.
\newblock \url{https://media.xiph.org/video/derf/}.

\bibitem{vimeo90k}
Tianfan Xue, Baian Chen, Jiajun Wu, Donglai Wei, and William~T Freeman.
\newblock Video enhancement with task-oriented flow.
\newblock {\em International Journal of Computer Vision (IJCV)},
  127(8):1106--1125, 2019.

\bibitem{yang2023asymmetrically}
Yibo Yang and Stephan Mandt.
\newblock Asymmetrically-powered neural image compression with shallow
  decoders.
\newblock {\em arXiv preprint arXiv:2304.06244}, 2023.

\bibitem{Zhu_Yang_Cohen_2022}
Yinhao Zhu, Yang Yang, and Taco Cohen.
\newblock Transformer-based transform coding.
\newblock In {\em International Conference on Learning Representations}, Mar
  2022.

\end{thebibliography}
}

\clearpage
\appendix

\section{Appendix}

\subsection{Commands for standard codecs}
\label{app:standard_codecs_commands}

The following commands are used to obtain compression results for standard codecs FFmpeg (x264 and x265) and HM.
For FFmpeg, we disable B-frames and use default settings otherwise.
We use HM-16.25 with default settings using the LowDelay-P config, for more details see \href{https://vcgit.hhi.fraunhofer.de/jvet/HM/-/tags/HM-16.25}{https://vcgit.hhi.fraunhofer.de/jvet/HM/-/tags/HM-16.25}.

\begin{small}
\begin{lstlisting}
# ffmpeg x264
ffmpeg -y -f rawvideo \ 
  -pix_fmt yuv420p \
  -s:v <width>x<height> \ 
  -i <input.yuv> \
  -r <framerate> \ 
  -c:v libx264 \
  -preset <preset> \  
  -crf <crf> \ 
  -x264-params bframes=0 \
  <output>

# ffmpeg x265
ffmpeg -y -f rawvideo \ 
  -pix_fmt yuv420p \
  -s:v <width>x<height> \ 
  -i <input.yuv> \ 
  -r <framerate> \ 
  -c:v libx265 \
  -preset <preset> \  
  -crf <crf> \ 
  -x265-params bframes=0 \
  <output>

# HM-16.25 LowDelayP
./bin/TAppEncoderStatic -c \ 
  ./cfg/encoder_lowdelay_P_main.cfg \
  -i <input.yuv> \
  --InputBitDepth=8 \ 
  -wdt <width> \
  -hgt <height> \
  -fr <framerate> \
  -f <numframes> \ 
  -q <qp> \
  -o <output>
\end{lstlisting}
\end{small}

\newpage

\subsection{Source Data}
Per-video and per-color channel benchmark results are included in a csv file in the Supplementary Materials, examples of videos decoded with our codec can be viewed at \url{https://www.youtube.com/watch?v=jXH6utaZirU}.

\subsection{Additional Results}
\begin{figure}[b]
    \centering
    \includegraphics[width=0.9\linewidth]{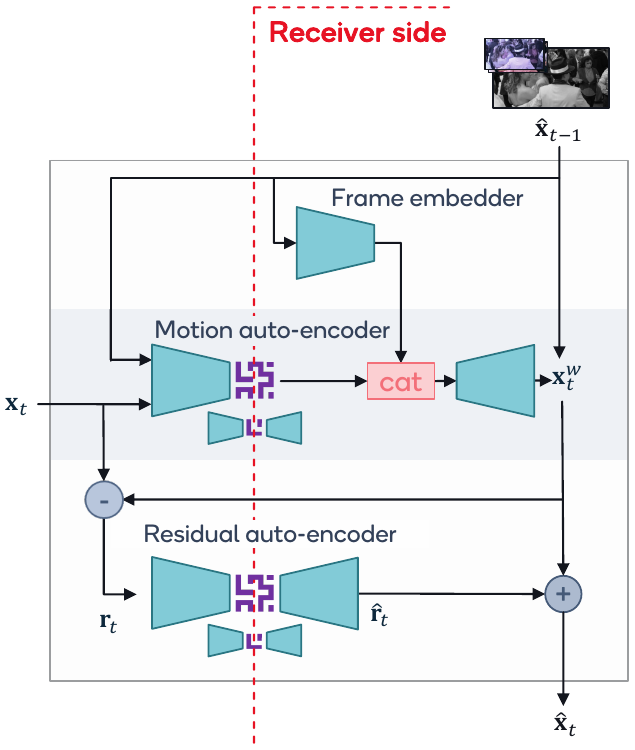}
    \caption{Model architecture of our flow-agnostic model. \protect\footnotemark}
    \label{fig:model_arch_flow_agnostic}
\end{figure}

\begin{figure*}
    \centering
    \includegraphics[width=0.95\linewidth]{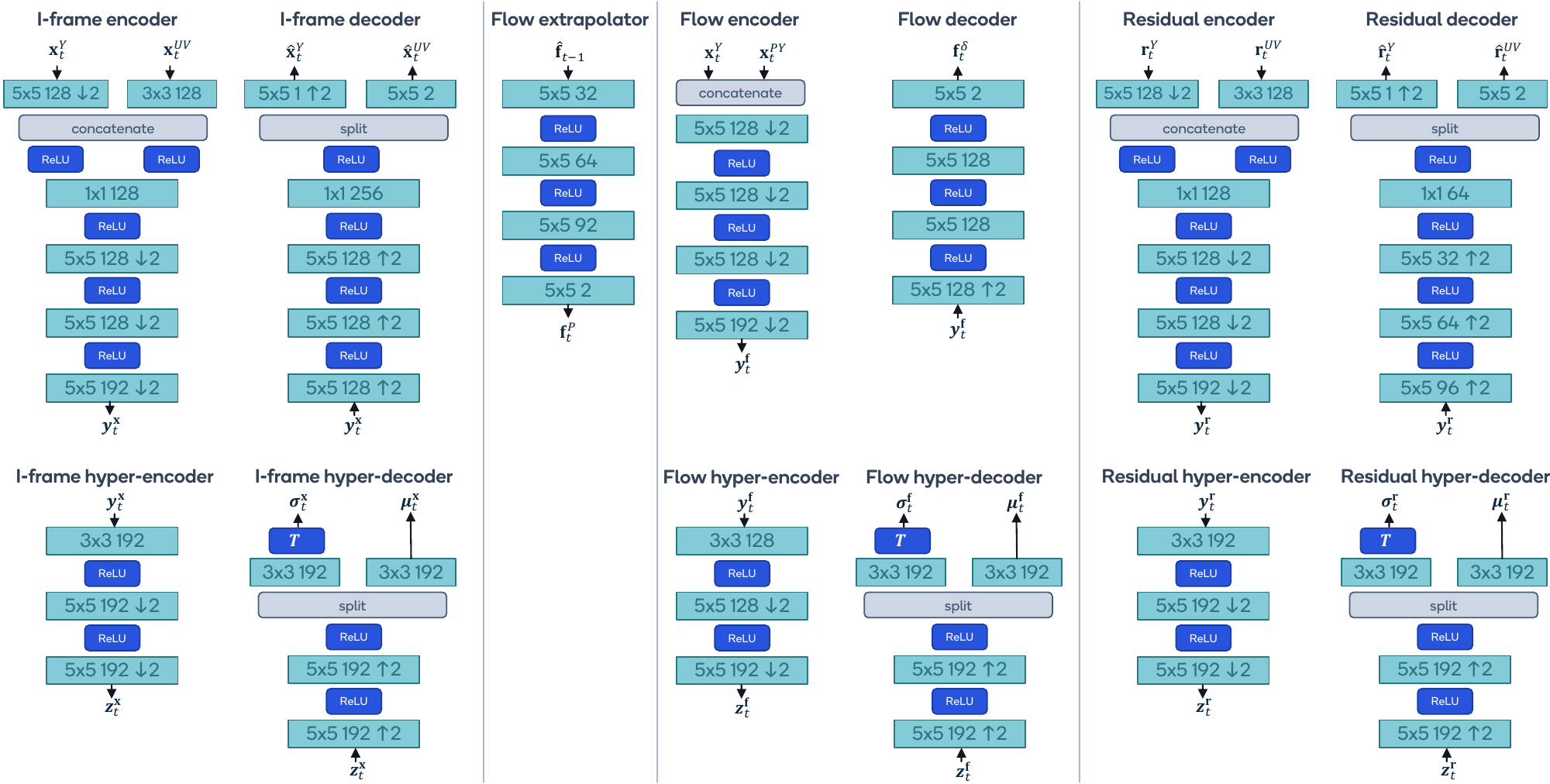}
    \caption{Model architecture for the neural networks inside of our P-frame model. Convolutional layers are displayed as $k\times k \ c$ where $k$ refers to kernel size and $c$ refers to the number of output channels. Convolutions with stride $s$ are indicated by $\downarrow s$ and transposed convolutions with stride $s$ are shown as $\uparrow s$.}
    \label{fig:full_model_architecture}
\end{figure*}

\begin{figure*}
    \centering
    \includegraphics[width=0.93\textwidth]{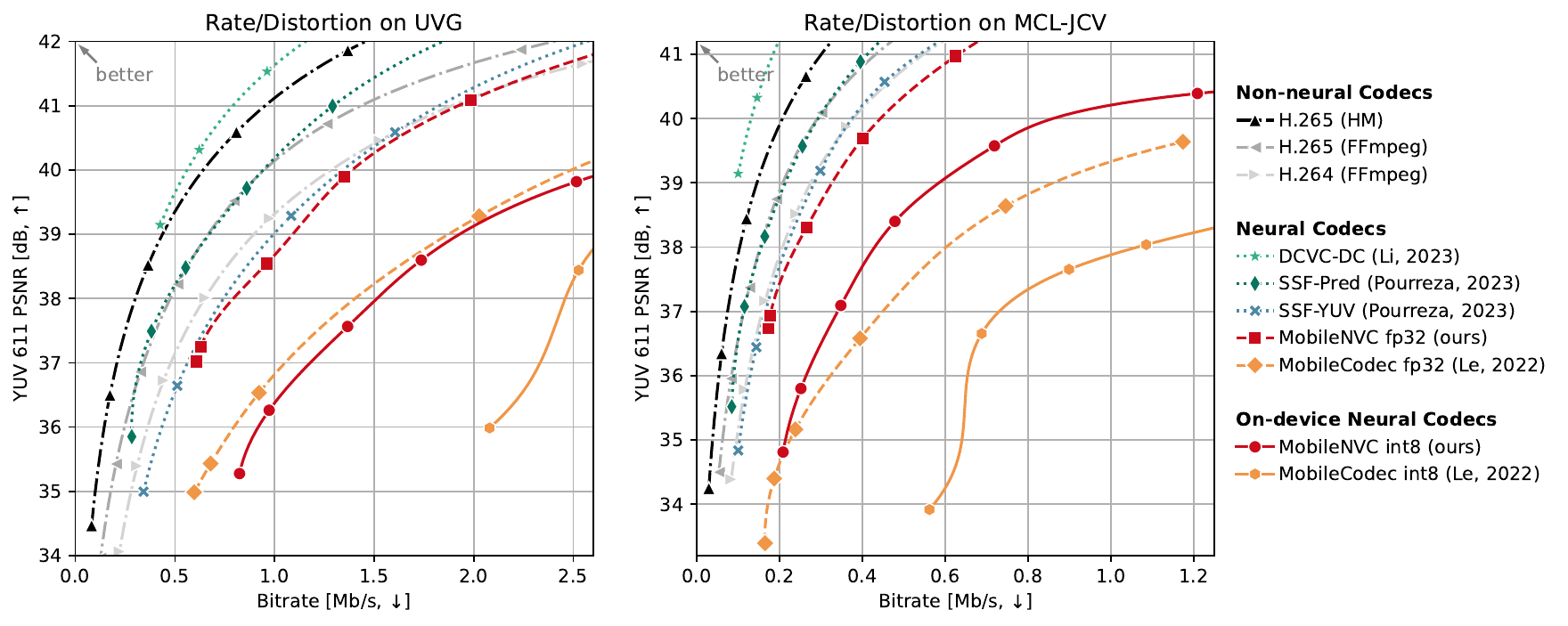}
    \caption{Rate-distortion performance of all models on UVG and MCL-JCV.}
    \label{fig:RD_results_uvg_mcl}
\end{figure*}

\begin{figure*}
    \centering
    \includegraphics[width=\textwidth]{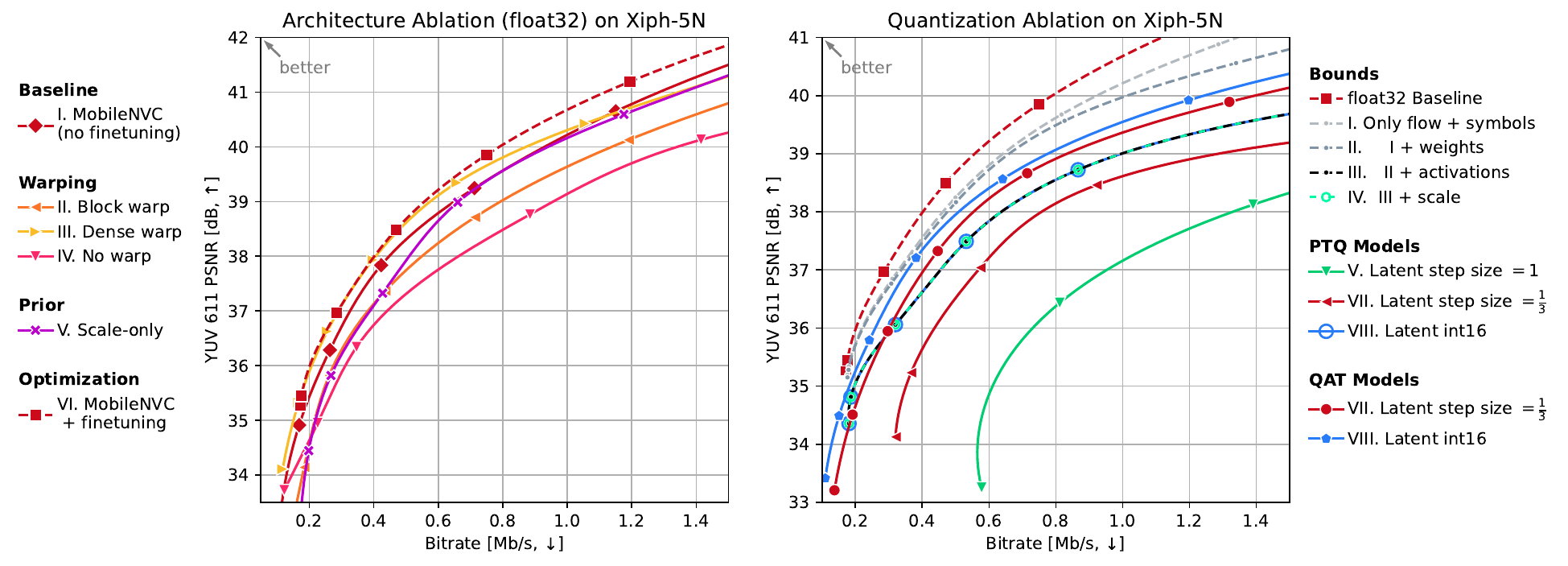}
    \caption{Model ablation (Left) and Quantization Ablation (Right). The models in these plots are described in more detail in \Cref{tab:architecture_ablation} and \ref{tab:quantization_ablation} respectively.}
    \label{fig:RD_model_ablations}
\end{figure*}

Additional results are shown on the following pages. \Cref{tab:training_stages} lists the hyperparameters used in the various training stages of our model. The full model architecture is detailed in \Cref{fig:full_model_architecture}. 
\Cref{fig:RD_model_ablations} shows the RD performance for the models discussed in the model and quantization ablation in the main text of our paper and \Cref{fig:model_arch_flow_agnostic} details the pipeline for our flow-agnostic model in this ablation. Finally, \Cref{fig:RD_results_uvg_mcl}  shows the benchmark of our model and various baselines on the UVG and MCL-JVC datasets.

\footnotetext{Image data from Tango video from
Netflix Tango in Netflix El Fuente. Video produced by Netflix, with CC BY-NC-ND 4.0 license: \url{https://media.xiph.org/video/derf/ElFuente/Netflix_Tango_Copyright.txt}}

\begin{table*}
    \footnotesize
    \centering
        \begin{tabular}{llllll}
    \toprule
    \textbf{}                                  & \textbf{}                        & \textbf{Stage 1}  & \textbf{Stage 2}    & \textbf{Stage 3} & \textbf{Stage 4}     \\
    \textbf{}                                  & \textbf{}                        & \textbf{Training} & \textbf{finetuning} & \textbf{PTQ}     & \textbf{QAT}         \\
    \midrule
    \textbf{Data Size}
    & {batchsize}               & 8                 & 16                  & 2                & 16                   \\
                                               & {gop}                     & 4                 & 7                   & 3                & 4                    \\
                                               & {crop size}               & 256x256           & 256x384             & 256x256          & 256x256              
     \\[10pt]
    \textbf{Loss Multipliers}
    & {$\beta$ I-frame}             & $\beta$         &  $\beta$            &  $\beta$          & $\beta$                \\
                                               & {$\beta$ P-frame}             & $2\beta$            & $2\beta$              & -                & $2\beta$               \\
                                               & {P-frame loss modulation}  & $\tau=1$ (no modulation)            & $\tau=1.2$                 & -                & $\tau=1.2$                  \\
                                               & {predicted flow $\fvec^P$}    & $\lambda=0.1$               & $\lambda=0$                   & -                & $\lambda=0$                    \\
                                               & {reconstructed flow $\hat{\fvec}$}   & $\lambda=0.1$               & $\lambda=0$                   & -                & $\lambda=0$           
 \\[10pt]
    \textbf{Optim}
    & {lr}                      & 1e-04          & 5e-05            & -                & 5e-07             \\
                                               & {lr schedule}             & -                 & -                   & -                & cosine decay to 1e-9 
   \\[10pt]
    \textbf{Quantization}                      & {datatype}                        & float32           & float32             & int8             & int8 (STE)           
    \\[10pt]
    \textbf{Training Time}
    & {steps}                   & 1M                & 250k                & 30               & 100k                 \\
                                               & {walltime}                & $\sim$ 4 days                  & $\sim$ 3 days                    & $\sim$ 2 minutes                 & $\sim$ 1 day                    \\
    \bottomrule
    \end{tabular}
    \caption{Different training stages and their corresponding hyperparameters. \\ We train a model for each value of  $\beta \in \{ 0.0001, 0.0002, 0.0004, 0.0008, 0.0016, 0.0032, 0.0064\}$}
    \label{tab:training_stages}
\end{table*}

\begin{table*}
    \footnotesize
    \centering
    \begin{tabular}{llrrrrlrrrr}
    \toprule
                                       &                             & \multicolumn{4}{l}{\textbf{Parameters [M, $\downarrow$]}}                              &  & \multicolumn{4}{l}{\textbf{kMACs/px [$\downarrow$]}}                                    \\
    \cmidrule{3-6}\cmidrule{8-11}
                                       &                             & \textbf{MobileNVC} & \textbf{MobileCodec} & \textbf{SSF} & \textbf{SSF-Pred} &  & \textbf{MobileNVC} & \textbf{MobileCodec} & \textbf{SSF} & \textbf{SSF-Pred} \\
    \midrule
    \multirow[t]{6}{*}{\textbf{Sender}}   & \textbf{I-frame AE}         & 5.66               & 6.68                & 9.47             & 9.47                  &  & 116.11              & 211.60                & 118.01             & 118.01                  \\
    \cmidrule{2-11}
                                       & Motion pred.   &  0.21              & -                    & -            & 0.75                  &  & 1.66               & -                    & -            & 6.44                  \\
                                       & Motion AE          & 5.59               & 11.63                & 9.48             & 10.10                  &  & 28.34               & 175.60                & 118.70             & 124.14                  \\
                                       & Residual pred. & -              & -                    & -            &  0.75                 &  & -              & -                    & -            & 6.32                  \\
                                       & Residual AE        & 6.82               & 6.57                & 10.09             & 10.09                  &  & 36.59               & 183.60                & 123.45             & 123.45                  \\
                                       & \textbf{Pframe total}       & 12.42              & 18.20               & 19.57             & 21.69                  &  & 64.93               & 359.20                &           242.15   & 260.35                  \\
\midrule
    \multirow[t]{6}{*}{\textbf{Receiver}} & \textbf{I-frame AE}         & 2.94               & 2.94                & 5.82             & 5.82                  &  & 93.39               & 130.90                & 94.64             & 94.64                  \\
    \cmidrule{2-11}
                                       & Motion pred.   & 0.21               & -                    & -            & 0.75                  &  & 1.66               & -                    & -            & 6.44                  \\
                                       & Motion AE          & 2.91               & 5.98                 & 5.83             & 6.45                  &  & 9.5               & 156.20                & 94.64             & 100.08                  \\
                                       & Residual pred. & -              & -                    & -            & 0.75                  &  & -              & -                    & -            & 6.32                  \\
                                       & Residual AE        & 3.18               & 2.75                & 6.44             & 6.44                  &  & 13.36               & 102.90                & 100.08             & 100.08                  \\
                                       & \textbf{Pframe total}       & 6.30              & 8.65                &  12.27            & 14.39                  &  & 24.52               & 259.10                & 194.72             & 212.92                 \\
\bottomrule                                       
    \end{tabular}
    \caption{Model complexity per subnetwork for $1080 \times 1920$ YUV420 input. 
    AE refers to (hypper-prior) autoencoder components and pred. refers to predictor models. Models are: MobileNVC (ours), MobileCodec \cite{le2022mobilecodec}, SSF \cite{agustsson2020ssf}, and SSF-Pred \cite{pourreza2022boosting}.
    }
    \label{tab:model_complexity_per_subnetwork}
\end{table*}

\clearpage

\onecolumn

\subsection{Warping Samples}

\begin{table*}[b!]
\centering
\setlength\tabcolsep{0pt}
\begin{tabular}{lcccc}
             & \textbf{III. Dense Warp} \vspace{1mm}                                      & \textbf{II. Block Warp}                                       & \textbf{I. Block-Overlap}                   & \textbf{IV. No Warp}
\\
\rotatebox[]{90}{\makecell{\textbf{Transmitted Flow} \\ $\hat{\fvec}_t$}}
& \includegraphics[height=3.8cm, valign=c]{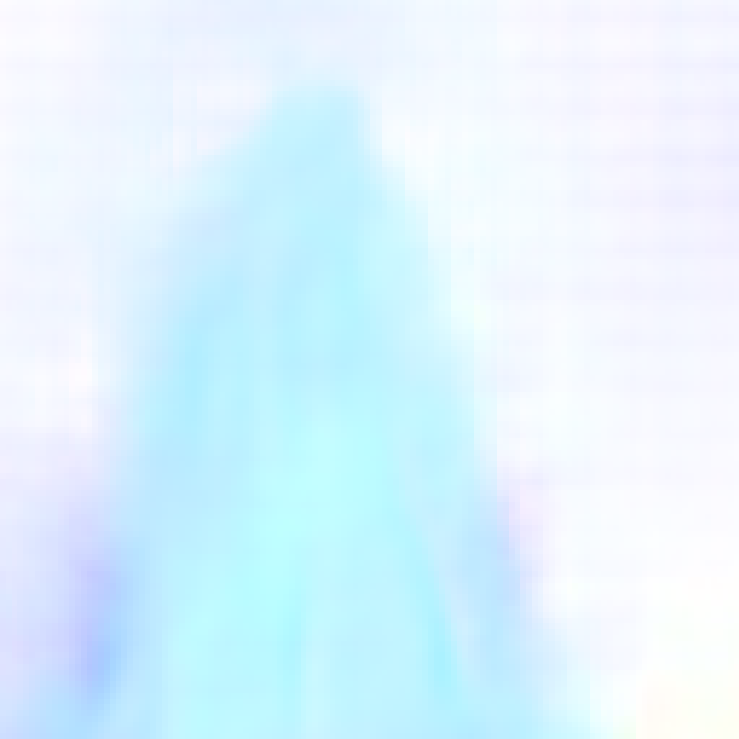}
& \includegraphics[height=3.8cm, valign=c]{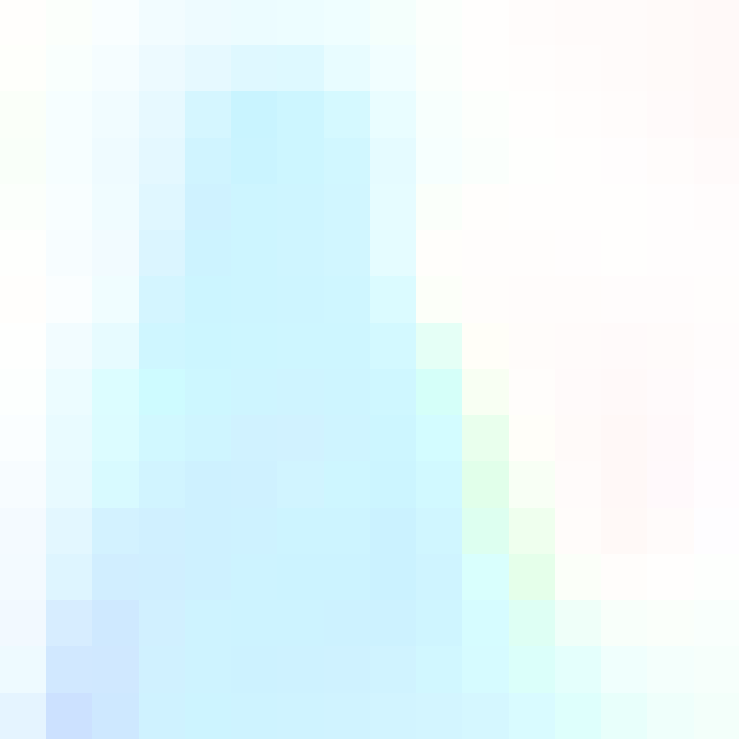}
& \includegraphics[height=3.8cm, valign=c]{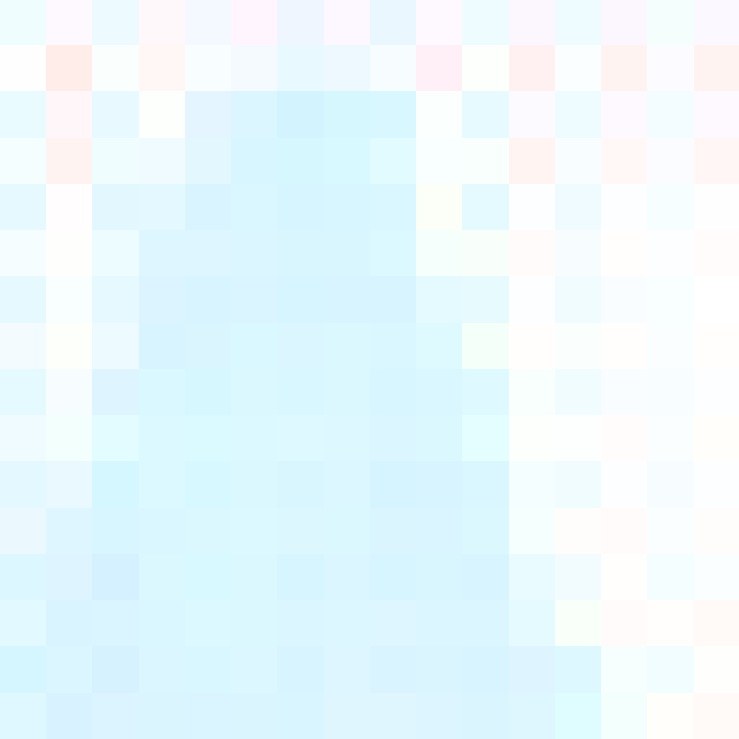}
& 
\\
\rotatebox[]{90}{\makecell{\textbf{Warped Frame} \\ $\operatorname{warp}(\xvec_{t-1}, \hat{\fvec}_t)$}} \hspace{1mm}
& \includegraphics[height=3.8cm, valign=c]{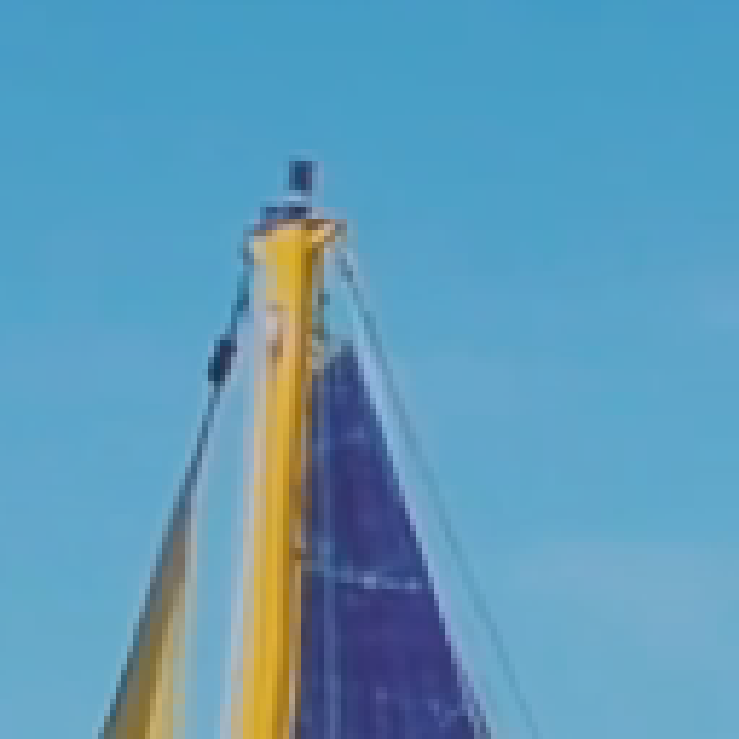}
& \includegraphics[height=3.8cm, valign=c]{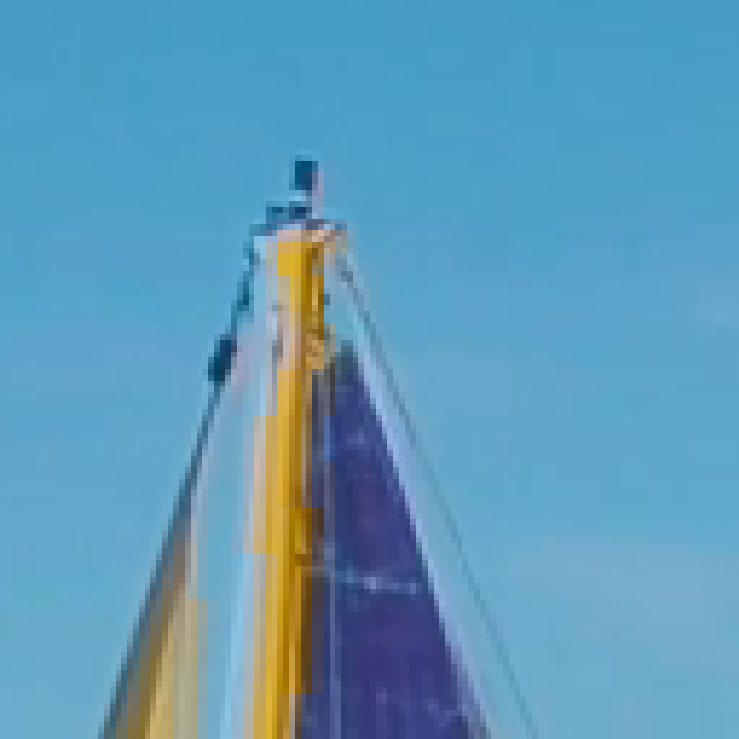}
& \includegraphics[height=3.8cm, valign=c]{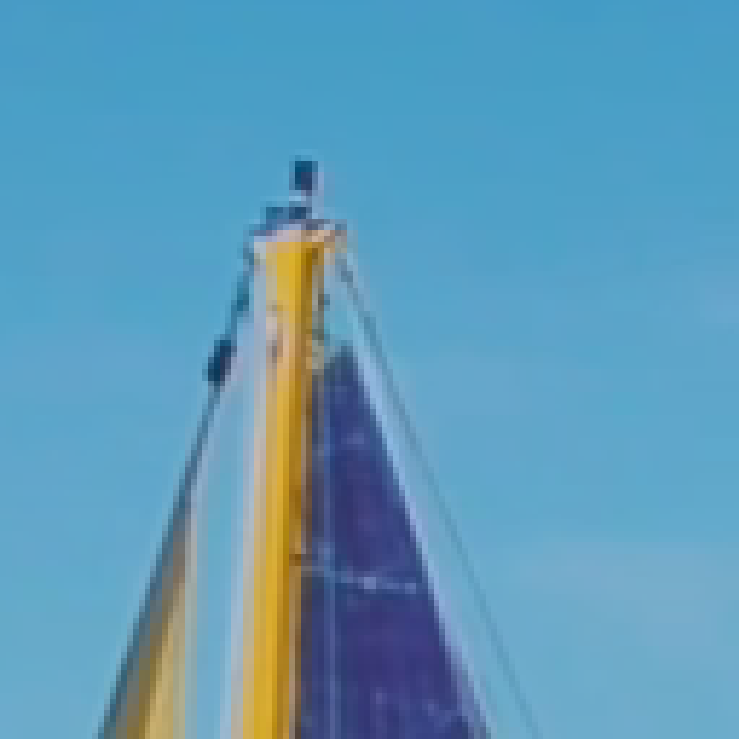}
& \includegraphics[height=3.8cm, valign=c]{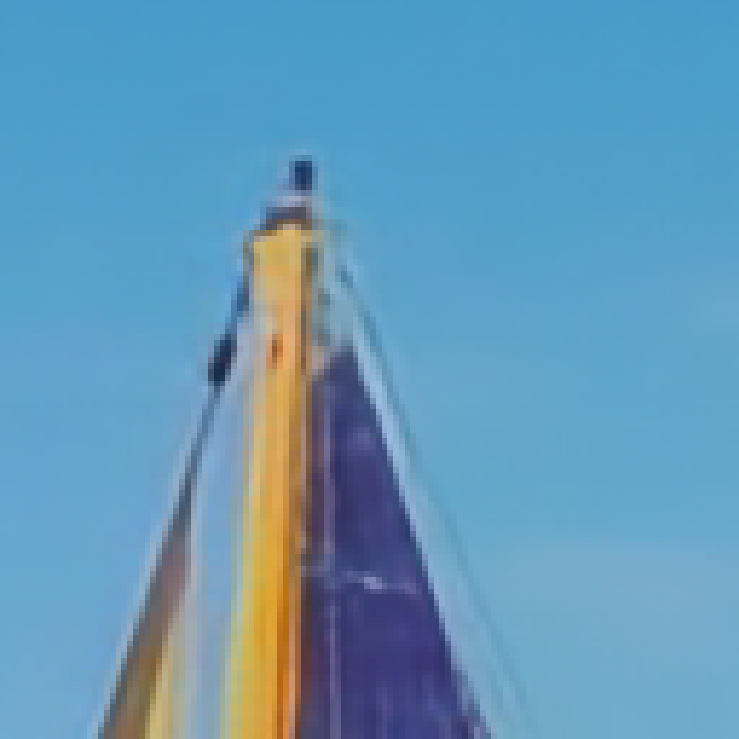}
\\
\rotatebox[]{90}{\makecell{\textbf{Warping Residual} \\ $\xvec^W_t - \xvec_t$}}
& \includegraphics[height=3.8cm, valign=c]{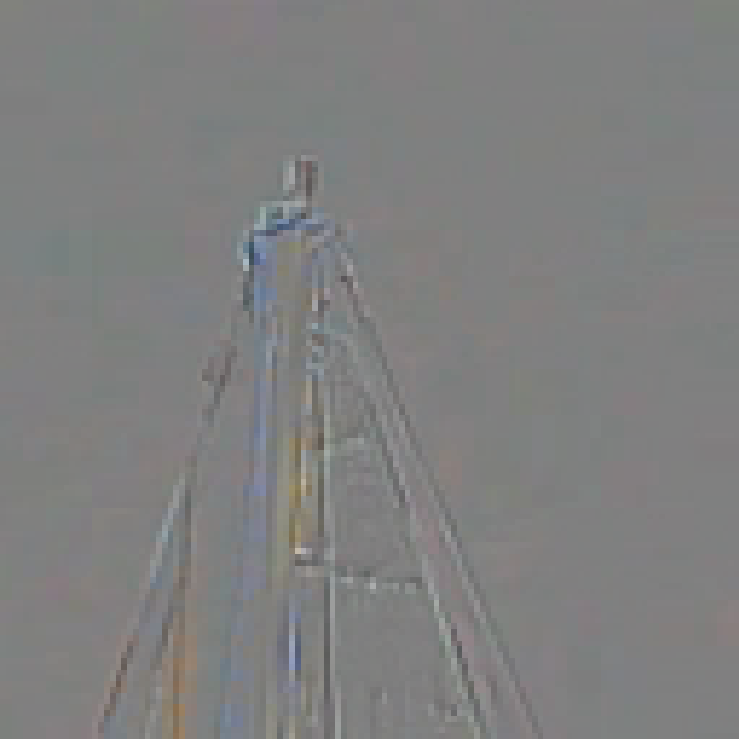}
& \includegraphics[height=3.8cm, valign=c]{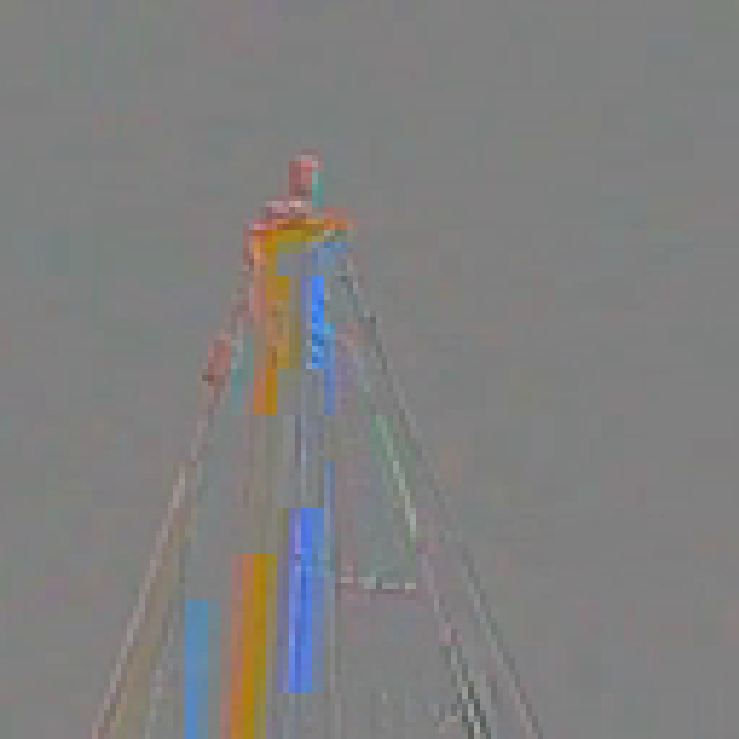}
& \includegraphics[height=3.8cm, valign=c]{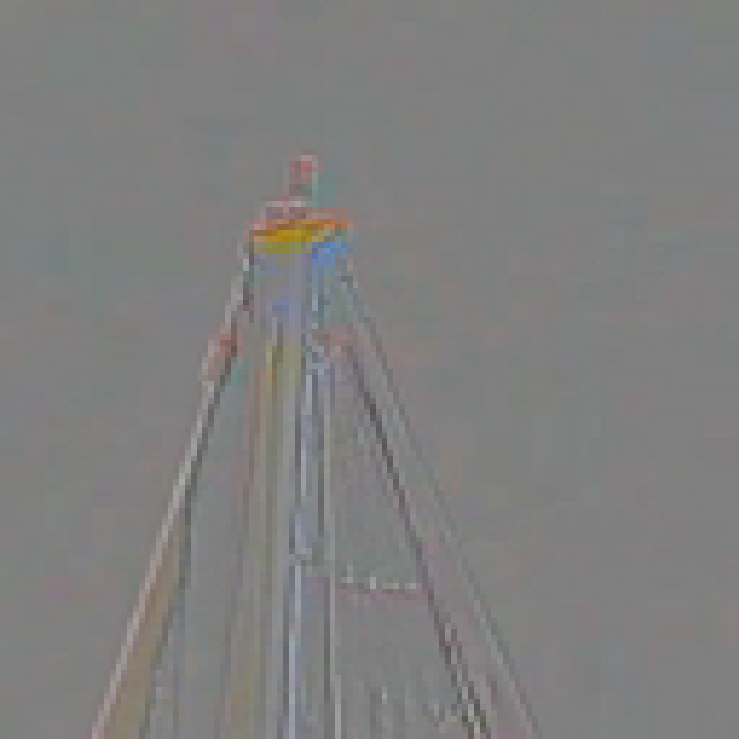}
& \includegraphics[height=3.8cm, valign=c]{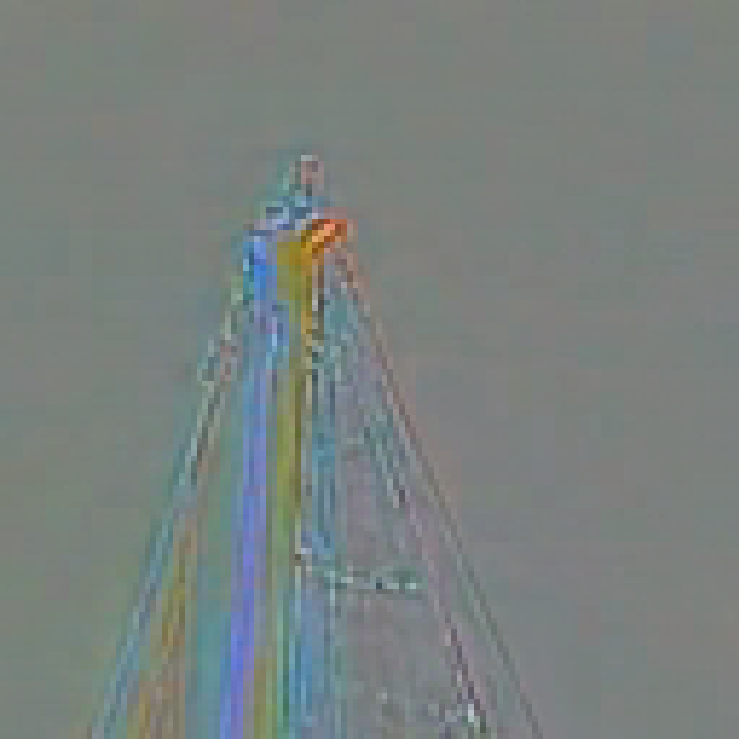}
\end{tabular}

\twocolumn
\onecolumn
\captionsetup{justification=centering}
\caption{Visualization of the output of different warping strategies. Numerals respond to the models in \cref{tab:architecture_ablation}. \\ Datapoint obtained from \url{https://www.pexels.com/video/sky-blue-boat-sailing-4602958}. Crop location: \\ \includegraphics[width=0.41\linewidth]{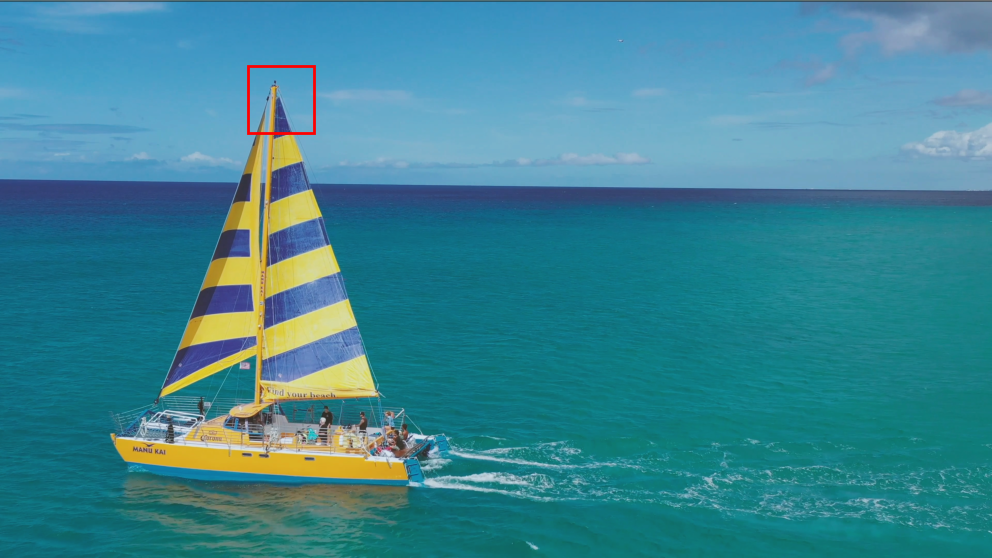}}
\label{tab:warping_samples}

\twocolumn
\end{table*}

The effect of the different warping strategies is shown in \cref{tab:warping_samples}. In this experiment, we compress a single P-frame as usual, but instead of conditioning the model on the previously reconstructed frame $\hat{\xvec}_{t-1}$ we use the previous groundtruth frame $\xvec_{t-1}$. Doing so allows us focus on the differences in warping only.

The frame warped with dense warping (III) does not show any clear artifacts. When we use block-warp (II) instead, we see discontinuities where the edges of the object to warp do not align with the blocks (i.e. notice the "gaps" in the yellow mast pole). We see that for the Block-Overlap Warp (I) these artifacts have disappeared. Finally, when we do not use warping but deploy a conditional model (IV) instead, the predicted frame becomes a lot less crisp.

Note that a checkerboard-like pattern can be seen in the flow for our Block-Overlap warping model. This pattern arises as the network learns to exploit the merging of neighboring blocks in blending kernel for a better final result.

\end{document}